\documentclass[options,onecolumn]{mn2e}
\usepackage{epsfig}
\newif\ifAMStwofonts

\newcommand{\mt}[1]{\mbox{$\mathbfss{#1}$}}
\newcommand{\wigner}[6]
{{
\left(
\begin{array}{ccc}
#1 & #2 & #3 \\
#4 & #5 & #6 \\
\end{array}
\right)
}}

\newcommand{\wigo}[3]{\wigner{\ell_{#1}}{\ell'_{#2}}{\ell''_{#3}}{0}{0}{0}}
\newcommand{\VEV}[1]{\langle#1\rangle}

\title[CMB Power Spectrum Estimation of Temperature and Polarisation]
{Fast CMB Power Spectrum Estimation of Temperature and Polarisation with Gabor Transforms}
  
\author[Frode K. Hansen and Krzysztof M. G\'orski]
  {{Frode K. Hansen$^1$, \thanks{E-mail: frodekh@roma2.infn.it}} and {Krzysztof M. G\'orski$^{2,3}$, \thanks{E-mail: kgorski@eso.org}}\\
$1$ Dipartimento di Fisica, Universit\`a di Roma `Tor Vergata', Via della Ricerca Scientifica 1, I-00133 Roma, Italy\\
$2$ ESO, Karl-Schwarzschild-Str.2, 85748 Garching bei M\"unchen, Germany\\
$^3$ Warsaw University Observatory, Aleje Ujazdowskie 4,00-478 Warszawa, Poland\\}
\pagerange{\pageref{firstpage}--\pageref{lastpage}}
\pubyear{2002}

\begin{document}

\label{firstpage}

\maketitle

\begin{abstract}
We extend the analysis of Gabor transforms on a Cosmic Microwave Background (CMB) temperature map \cite{gabortrans} to polarisation. We study the temperature and polarisation power spectra on the cut sky, the so-called pseudo power spectra. The transformation kernels relating the full-sky polarisation power spectra and the polarisation pseudo power spectra are found to be similar to the kernel for the temperature power spectrum. This fact is used to construct a fast power spectrum estimation algorithm using the pseudo power spectrum of temperature and polarisation as data vectors in a maximum likelihood approach. Using the pseudo power spectra as input to the likelihood analysis solves the problem of having to invert huge matrices which makes the standard likelihood approach infeasible.  
\end{abstract}

\section{introduction}

Most theories of the early universe predict the temperature and polarisation fluctuations of the CMB to be Gaussian distributed. In such models, the angular temperature and polarisation power spectra contain all the information about the cosmological parameters which one can determine from observations of the CMB sky. As several combinations of the cosmological parameters can give rise to similar temperature power spectra, estimating the polarisation power spectra will break the degeneracy and will be of great importance for accurate estimation of cosmological parameters. Also the error bars on these parameters can be reduced by exploiting the extra information present in the CMB
polarisation power spectra.\\

Much effort has been made recently in order to find methods to analyse the CMB
temperature power spectrum \cite{gabortrans,OhSpergelHinshaw,pseudo,ringtorus1,ringtorus2,bond,BJK,bartlett,tegmark,dore,szapudi,master,amad}. There has been very few publications confronting the even harder task of estimating the
polarisation power spectra. The
framework for analysing the polarisation power spectra has been set in
\cite{pol1,pol2} but these papers only describe the full likelihood
method which is far too time consuming also when only considering the
temperature power spectrum. In \cite{tegmark0} a quadratic polarisation power spectrum estimation method was introduced, similar to the one presented in \cite{tegmark} for temperature only.\\

In this paper we will extend the method of using the pseudo power
spectrum as input to a likelihood estimation procedure of the power spectrum as described in \cite{gabortrans} (from now on called HGH). We
will include the $E$ and $C$ mode polarisation pseudo power spectra
in the data vector and use techniques similar to those described in HGH to estimate the power spectra. This can be done
because the kernels that connect the full sky polarisation power
spectra with the polarisation pseudo power spectra on an apodised sky are
similar to the kernel for the temperature power spectrum. In the first
part of this paper we will derive the formulae for these kernels and
for the polarisation pseudo power spectra and discuss their
shapes. Then in the second part this will be used for likelihood
estimation.\\

In this paper the $B$ component polarisation will mostly be
neglected. The $B$ polarisation power spectrum is expected to be very
small and will hardly be detectable by the upcoming $MAP$ or $Planck$
satellite experiments \cite{jaffe0}. Also the $E$ and $B$ components of polarisation mix on
the cut sky as will be discussed in this paper, making the $B$
polarisation pseudo spectrum to be dominated by the $E$ component \cite{sepeb1,sepeb2,tegmark0,sepeb3}.

\section{The Gabor Transformation}
\label{sect:polpowspec}

In this paper we will use the polarisation power spectra as they are defined in \cite{pol1}. There are three polarisation power spectra, $C^E_\ell$, $C^B_\ell$ and $C^C_\ell$. These are defined as
\begin{eqnarray}
C^E_\ell&=&\sum_m\frac{a^E_{\ell m}a^{E*}_{\ell m}}{2\ell+1},\\
C^B_\ell&=&\sum_m\frac{a^B_{\ell m}a^{B*}_{\ell m}}{2\ell+1},\\
C^C_\ell&=&\sum_m\frac{a^E_{\ell m}a^{T*}_{\ell m}}{2\ell+1},\\
\end{eqnarray}
where the $a_{\ell m}$ coefficients are given in Appendix (\ref{app:pseudopol}). In the following when we write the full sky power spectrum $C_\ell$ for temperature or polarisation, we will always mean the ensemble averaged power spectrum $\VEV{C_\ell}$.\\

As discussed in more detail in HGH, the Gabor transformation of a dataset \cite{gabor} is just the Fourier transformation of the dataset multiplied with a window function called the Gabor window. The window function can be used to cut out parts of the dataset in order to study only smaller segments of the dataset at a time. The window can be a top-hat window, just setting the unwanted parts of the dataset to zero. Another option is to use a function like a Gaussian to smooth the edges between the segment which one wants to study and the parts which are set to zero in order to avoid ringing in the Fourier spectrum. In HGH this formalism was extended to the sphere and used for CMB analysis. A disc on the CMB sky was cut out, using either a top-hat or a Gaussian window. The Gabor coefficients on this cut sky, called the pseudo power spectrum was expressed in terms of the full-sky power spectrum and the kernel connecting the full-sky and the cut-sky power spectrum was studied. The aim of this first section is to extend this to polarisation.\\

When multiplying the polarisation map with a Gabor window $G(\mathbf{n},\mathbf{n}_0)$ which is axissymmetric about the point $\mathbf{n}_0$, the pseudo spectra on this apodised CMB sky can be written in terms of the full-sky spectra as (Appendix \ref{app:pseudopol})
\begin{eqnarray}
\label{eq:pcle}
\VEV{\tilde C_\ell^E}&=&\sum_{\ell'}C_{\ell'}^EK_2(\ell,\ell')+\sum_{\ell'}C_{\ell'}^BK_{-2}(\ell,\ell'),\\
\label{eq:pclb}
\VEV{\tilde C^B_\ell}&=&\sum_{\ell'}C_{\ell'}^B
K_2(\ell,\ell')+\sum_{\ell'}C_{\ell'}^EK_{-2}(\ell,\ell'),\\
\label{eq:pclc}
\VEV{\tilde C^C_\ell}&=&\sum_{\ell'}C_{\ell'}^C K_{20}(\ell,\ell'),\\
\end{eqnarray}
where the kernels $K_{\pm2}$ and $K_{20}$ are given in Appendix (\ref{app:pseudopol}). These kernels are expression in terms of the $h_2(\ell,\ell',m,m')$ function which is defined similar to the $h(\ell,\ell',m,m')$ function used to express the kernel for the temperature power spectrum in HGH.\\

As with the temperature kernels, the polarisation kernels can be
evaluated either using the analytical Wigner symbol expressions
(\ref{eq:kpm2}) and (\ref{eq:k0}) or faster by direct integration and
recursion of the $h_2(\ell,\ell',m,m')$ functions. The recursion for the
$h_2(\ell,\ell',m,m')$ functions in Appendix (\ref{app:recpol}) is one
of the major results in this paper. This is an extension of the
recursion for $h(\ell,\ell',m,m')$ derived in HGH.\\

In Appendix (\ref{app:pseudopol}) and (\ref{app:rotinv}) we show that the polarisation pseudo power spectra are rotationally invariant. For this reason we
will in the rest of the paper put the centre of the Gabor window on the
north pole. This makes the calculations easier while keeping the
generality of the results.\\

In this section we will study the Gabor kernel for a Gaussian and a top-hat Gabor window. The Gaussian is defined as 
\begin{eqnarray}
G(\theta)=&e^{-\theta^2/(2 \sigma^2)}&\theta\leq\theta_C,\\
G(\theta)=&0&\theta>\theta_C,
\end{eqnarray}
where we will use $5$ and $15$ degrees FWHM (corresponding to
$\sigma=2.12^\circ$ and $\sigma=6.38^\circ$) and a cut-off angle $\theta_C=3\sigma$. We will also be comparing with a top-hat window covering the same area on the sky ($\theta_C$ is the same). The top-hat window is defined as
\begin{eqnarray}
G(\theta)=&1&\theta\leq\theta_C,\\
G(\theta)=&0&\theta>\theta_C.
\end{eqnarray}

\begin{figure}
\begin{center}
\leavevmode
\psfig {file=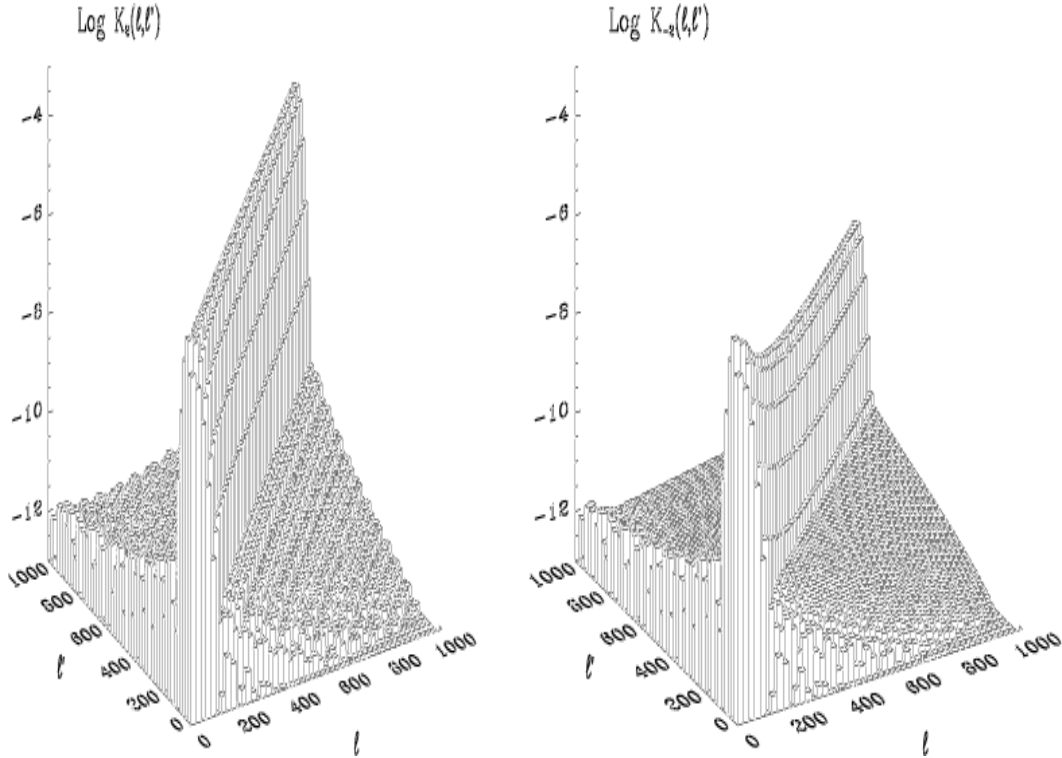,height=10cm,width=14cm}
\caption{The kernels \protect{$K_2(\ell,\ell')$} (left plot) and
\protect{$K_{-2}(\ell,\ell')$} (right plot) connecting full and cut sky
polarisation power spectra \protect{$C^E_\ell$} and
\protect{$C^B_\ell$}. The left kernel is the one which takes full sky
$C^E_\ell$ into cut sky $\tilde C^E_\ell$ and full sky $C^B_\ell$ into
cut sky $\tilde C^B_\ell$. The right kernel is the one which mixes the
two giving contributions from full sky $C^E_\ell$ in cut sky $\tilde
C^B_\ell$ and vice versa.}  
\label{fig:k2g5}
\end{center}
\end{figure}

\begin{figure}
\begin{center}
\leavevmode
\psfig {file=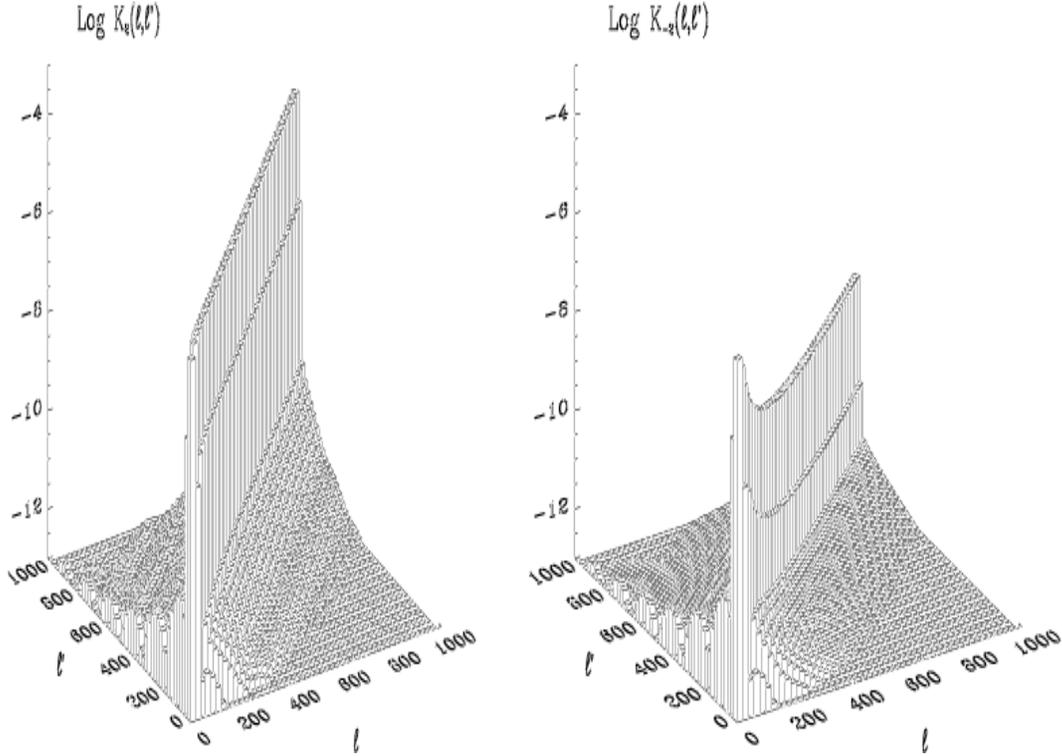,height=10cm,width=14cm}
\caption{The same as figure (\ref{fig:k2g5}) for a $15$ degree FWHM
Gaussian Gabor window.}  
\label{fig:k2g15}
\end{center}
\end{figure}

Studying equation (\ref{eq:pcle}) and (\ref{eq:pclb}) one
sees that the $E$ and $B$ modes are mixing when only a portion of the
sky is observed. The kernel $K_2(\ell,\ell')$ is the kernel which
takes full sky $C^E_\ell$ modes to the pseudo coefficients $\tilde
C^E_\ell$ and similarly for $C^B_\ell$. The kernel
$K_{-2}(\ell,\ell')$ is the one which causes the mixing. In figure
(\ref{fig:k2g5}) and (\ref{fig:k2g15}) we have plotted the kernels
$K_2(\ell,\ell')$ together with $K_{-2}(\ell,\ell')$ for a $5$ and
$15$ degree FWHM Gaussian Gabor window with $\theta_C=3\sigma$. One
can see that the diagonal of the $K_2(\ell,\ell')$ kernel is about an order of
magnitude larger than the diagonal of the mixing kernel $K_{-2}(\ell,\ell')$. This
means that $C^E_\ell$ would dominate $\tilde C^E_\ell$ and $C^B_\ell$
would dominate $\tilde C^B_\ell$ provided that the two spectra
$C^E_\ell$ and $C^B_\ell$ were of the same order of magnitude. However as
discussed before, the $C^B_\ell$ are expected to be considerably smaller
than $C^E_\ell$ in most cosmological models. For the $E$ mode this is
not a problem as $C^E_\ell$ will then hardly be affected. The problem
is the $B$ mode which in this case will be dominated by
the $E$ mode.\\

The separation of $E$ and $B$ modes of polarisation on the cut sky was
already discussed in \cite{sepeb1,sepeb2,tegmark0,sepeb3}. We will in this paper assume that the $B$ mode polarisation component is neglible, but note that a possible extension of the power spectrum estimation method outlined here to $B$ mode polarisation would be to define
\begin{eqnarray}
a_{+,\ell m}&=&a_{E,\ell m}+ia_{B,\ell m}\\
a_{-,\ell m}&=&a_{E,\ell m}-ia_{B,\ell m},
\end{eqnarray}
with the corresponding pseudo quantities
\begin{eqnarray}
\label{eq:pclp}
\tilde a_{+,\ell
m}&=&\sum_{\ell'}a_{+,\ell'm}(H_2(\ell,\ell',m)+H_{-2}(\ell,\ell',m)),\\
\label{eq:pclm}
\tilde a_{-,\ell
m}&=&\sum_{\ell'}a_{-,\ell'm}(H_2(\ell,\ell',m)-H_{-2}(\ell,\ell',m)),\\
\end{eqnarray}
written in terms of the $H_2$ functions defined in Appendix (\ref{app:pseudopol}).
For the power spectra one gets,
\begin{eqnarray}
\VEV{C^+_\ell}&\equiv&\sum_m\frac{\VEV{a_{+,\ell m}a_{+,\ell m}^*}}{2\ell+1}=\VEV{C^E_{\ell m}}+\VEV{C^B_{\ell m}},\\
\VEV{C^-_\ell}&\equiv&\sum_m\frac{\VEV{a_{-,\ell m}a_{-,\ell m}^*}}{2\ell+1}=\VEV{C^E_{\ell m}}-\VEV{C^B_{\ell m}}.\\
\end{eqnarray}
To get the pseudo power spectra one can use equations (\ref{eq:pclp})
and (\ref{eq:pclm}) to get
\begin{eqnarray}
\VEV{\tilde C^+_\ell}&=&\sum_{\ell'}C^+_\ell K_+(\ell,\ell'),\\
\VEV{\tilde C^-_\ell}&=&\sum_{\ell'}C^-_\ell K_-(\ell,\ell'),
\end{eqnarray}
where the kernels can be written
\begin{eqnarray}
K_+(\ell,\ell')&=&\frac{1}{2\ell+1}\sum_m(H_2(\ell,\ell,m)+H_{-2}(\ell,\ell',m))^2,\\
K_-(\ell,\ell')&=&\frac{1}{2\ell+1}\sum_m(H_2(\ell,\ell,m)-H_{-2}(\ell,\ell',m))^2.
\end{eqnarray}

Separation of $E/B$ modes will not be discussed further in this paper as this is extensively treated in the references above.\\

\begin{figure}
\begin{center}
\leavevmode
\psfig {file=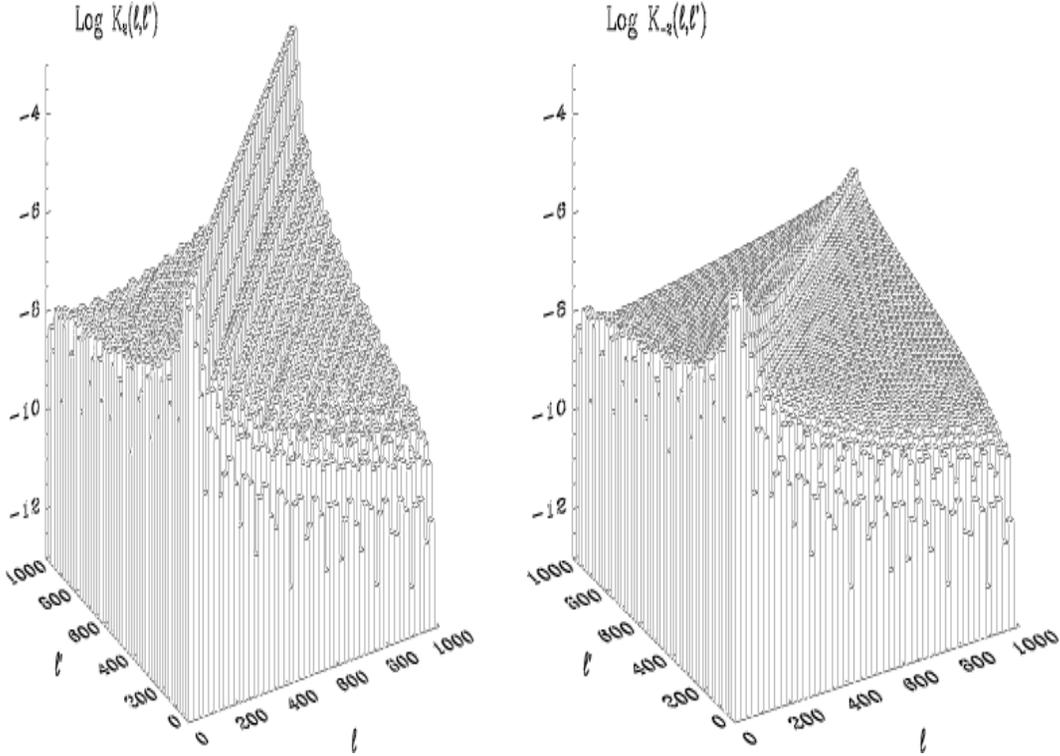,height=10cm,width=14cm}
\caption{The same as figure (\ref{fig:k2g5}) for a tophat Gabor window
covering the same area on the sky as the Gaussian window in figure (\ref{fig:k2g5})}  
\label{fig:k2th5}
\end{center}
\end{figure}

\begin{figure}
\begin{center}
\leavevmode
\psfig {file=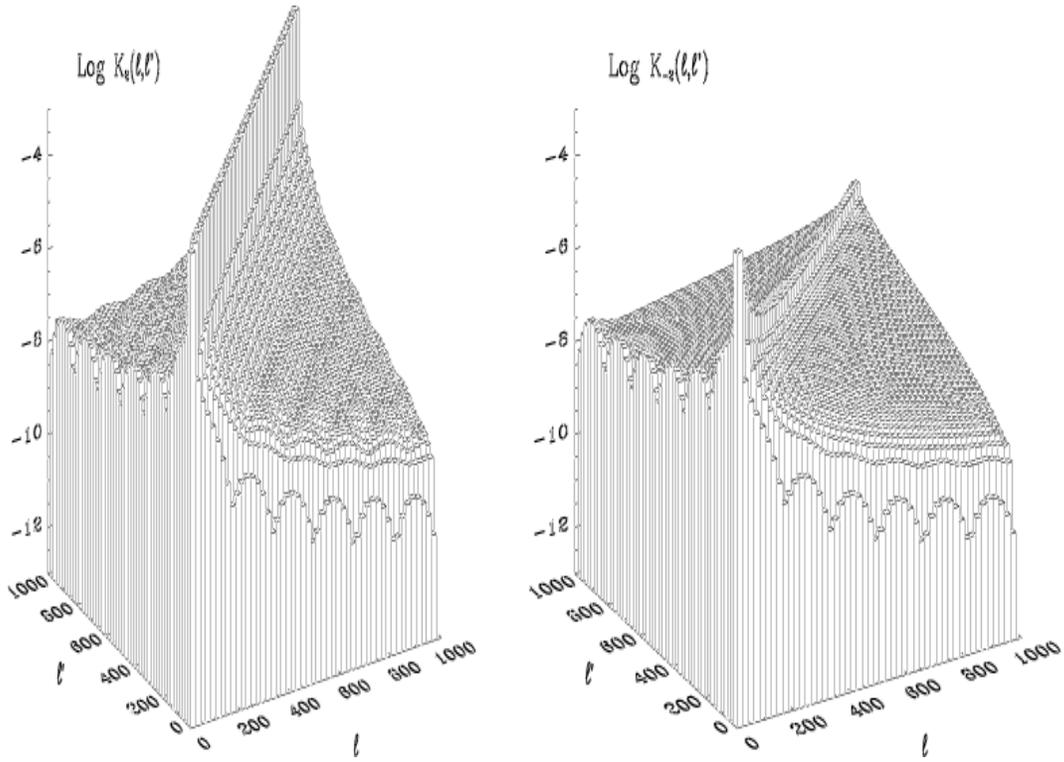,height=10cm,width=14cm}
 \caption{The same as figure (\ref{fig:k2g15}) for a tophat Gabor window
covering the same area on the sky as the Gaussian window in figure (\ref{fig:k2g15})}   
\label{fig:k2th15}
\end{center}
\end{figure}

In figure (\ref{fig:k2th5}) and (\ref{fig:k2th15}) we have plotted the
$K_2(\ell,\ell')$ and $K_{-2}(\ell,\ell')$ kernels for a tophat window
covering the same area on the sky as the Gaussian windows in figure
(\ref{fig:k2g5}) and (\ref{fig:k2g15}).

\begin{figure}
\begin{center}
\leavevmode
\psfig {file=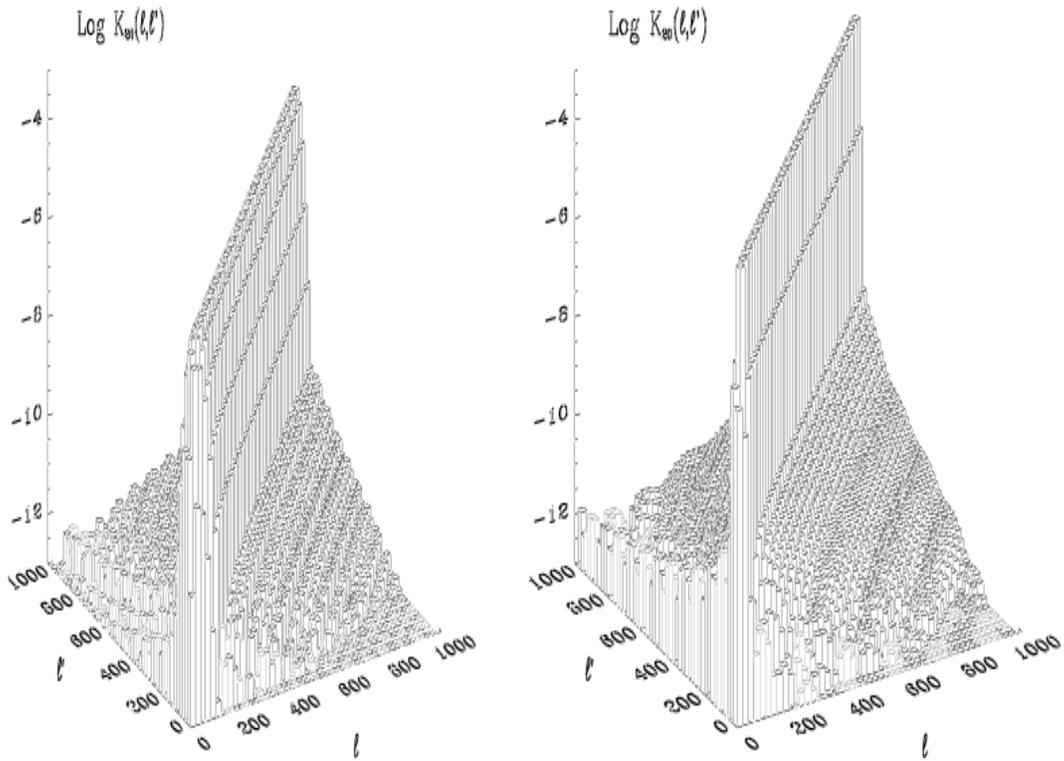,height=10cm,width=14cm}
\caption{The kernel \protect{$K_{20}(\ell,\ell')$} connecting the full sky cross polarisation spectrum
$C^C_\ell$ and the cut sky spectrum \protect{$\tilde C^C_\ell$} for a \protect{$5$} (left
plot) and \protect{$15$} (right plot) degree FWHM Gaussian Gabor
window. The negative elements have a brighter colour.}
\label{fig:k20g}
\end{center}
\end{figure}

\begin{figure}
\begin{center}
\leavevmode
\psfig {file=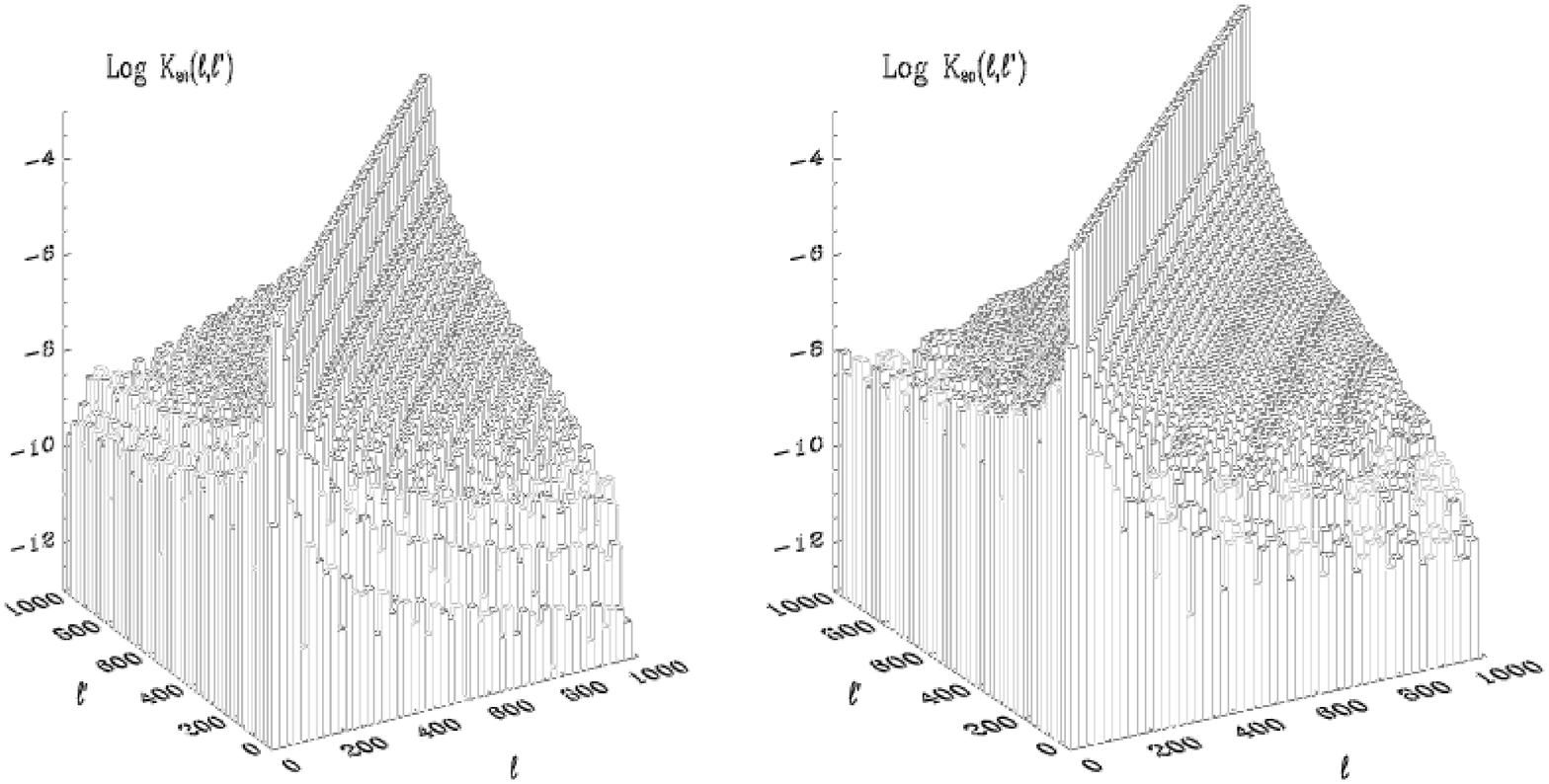,height=10cm,width=14cm}
\caption{Same as figure (\ref{fig:k20g}) for the corresponding tophat windows.}
\label{fig:k20th}
\end{center}
\end{figure}

The kernel $K_{20}(\ell,\ell')$ for the cross polarisation power
spectrum $C^C_\ell$ is shown in figure (\ref{fig:k20g}) for a $5$ and
$15$ degree Gaussian window and in figure (\ref{fig:k20th}) for the
corresponding tophat windows.

As for the temperature kernels, all the polarisation kernels show the
same behaviour when changing type and size of the window. When going
from smaller to larger windows, the diagonals get sharper. Also the
tophat kernels have more long range correlations than the Gaussian
kernels (note that all the plots have the same vertical scale and can
be compared directly).\\

\begin{figure}
\begin{center}
\leavevmode
\psfig {file=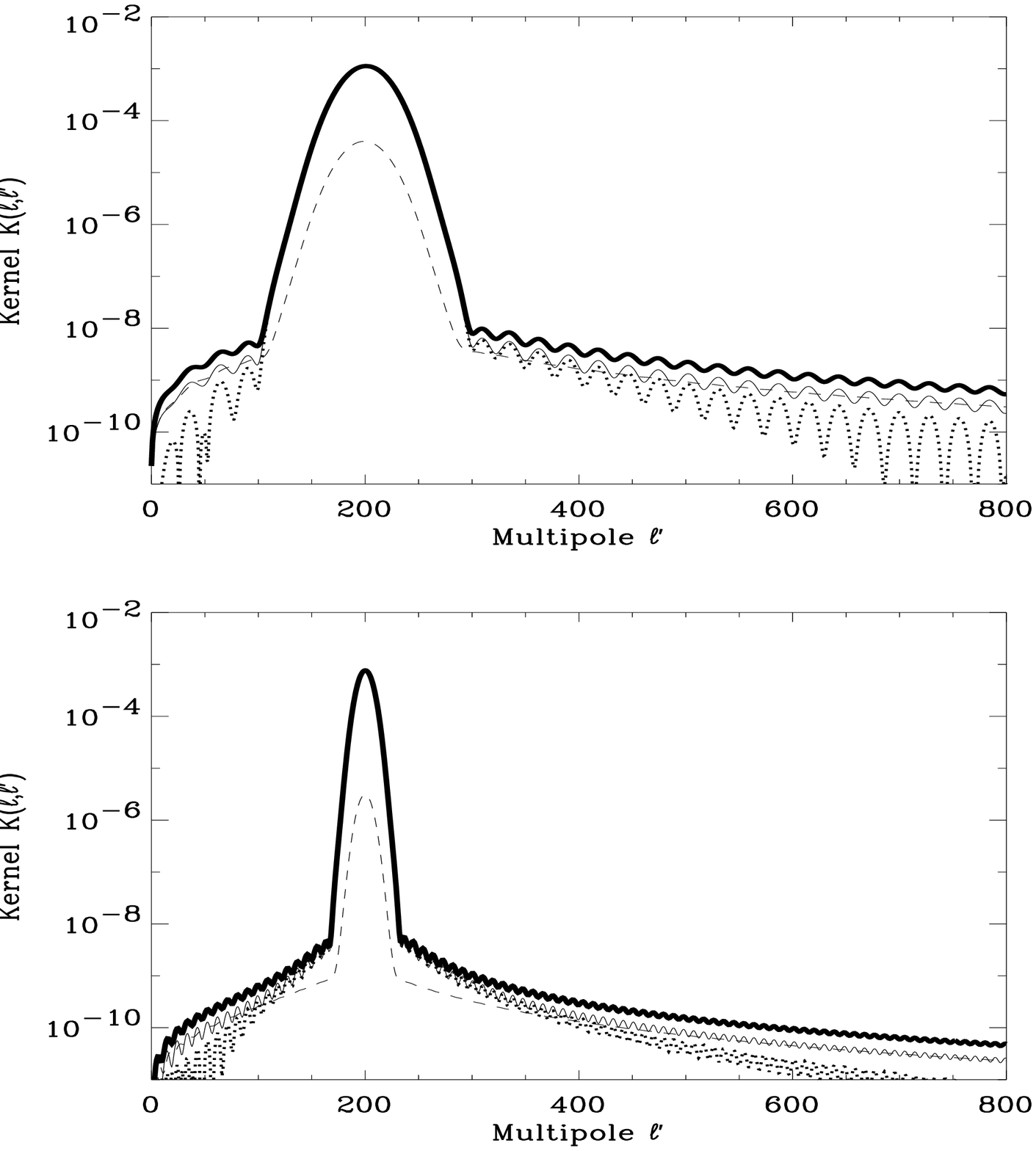,bbllx=0pt,bblly=150pt,bburx=596pt,bbury=842pt,height=12cm,width=14cm}
\caption{A slice at \protect{$\ell=200$} of the kernels combining
the full sky and cut sky power spectra. The thick solid line is the
kernel \protect{$K(\ell,\ell')$} for the temperature power spectrum, the thin solid line is
the kernel \protect{$K_2(\ell,\ell')$} for $E$ and $B$ mode
polarisation and the dotted line
is the kernel for the temperature-polarisation cross power spectrum
$K_{20}(\ell,\ell')$. All these kernels go together around the
diagonal. They only differ for the far off-diagonal elements. The
dashed line is the mixing kernel \protect{$K_{-2}(\ell,\ell')$} which
mixes the $E$ and $B$ mode polarisation power spectra on the cut
sky. This kernel is
lower than the other kernels. The upper plot is for a $5$ degree
Gaussian Gabor window and the lower plot for a $15$ degree FWHM
Gaussian window.}
\label{fig:polcutg}
\end{center}
\end{figure}

In figure (\ref{fig:polcutg}) we have plotted slices of the
different kernels at $\ell=200$ for comparison. The slices are made of the kernels for $5$ and $15$ degree FWHM Gaussian Gabor
windows. The first thing to
note is that the temperature kernel $K(\ell,\ell')$, the $E$ and $B$
kernel $K_2(\ell,\ell')$ and the temperature-polarisation cross
spectrum kernel $K_{20}(\ell,\ell')$ only differ for the far
off-diagonal elements. At the diagonal their shape and size are the
same. For this reason the relation shown in HGH
between the width of the kernel and the width of the size of the
window for the temperature power spectrum is also valid for polarisation. This is an
important result to be used for the likelihood estimation of the
polarisation power spectra in the next section. It shows that the
number of polarisation pseudo spectrum coefficients to be used in the
likelihood analysis should be the same as for the likelihood
estimation of the temperature power spectrum.\\

\begin{figure}
\begin{center}
\leavevmode
\psfig {file=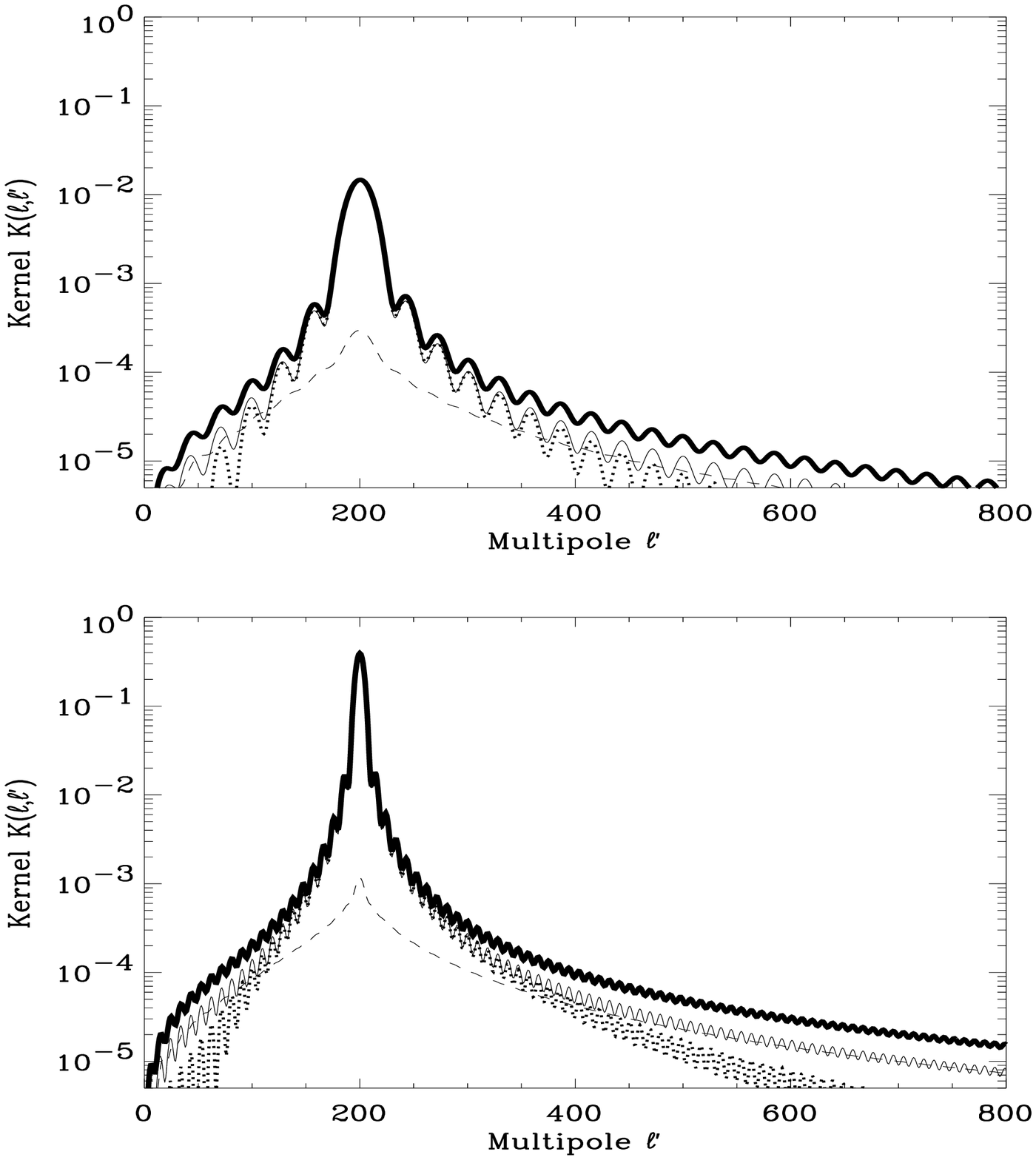,bbllx=0pt,bblly=150pt,bburx=596pt,bbury=842pt,height=12cm,width=14cm}
\caption{Same as figure (\ref{fig:polcutg}) but for tophat windows
covering the same area on the sky.}
\label{fig:polcutth}
\end{center}
\end{figure}

In figure (\ref{fig:polcutth})
a similar plot is shown for the corresponding tophat windows. The plot
shows that the conclusions made for the Gaussian windows are also
valid in this case. The shape and size of the three kernels are the
same around the diagonal. For this reason the results shown for the
temperature kernel that the tophat window has larger long range
correlations whereas the Gaussian window has large short range
correlations and therefore a wider kernel is also valid for polarisation.\\

The kernel $K_{-2}(\ell,\ell')$ which mixes the $E$ and $B$ modes on
the cut sky is plotted as a dashed line in figures (\ref{fig:polcutg})
and (\ref{fig:polcutth}). It is much smaller than the three other kernels and the shape
seems to differ as well. Note that the height of the mixing kernel
relative to the other kernels is lower for the $15$ degree window than
for the $5$ degree window. That the size of the mixing kernel relative
to the other kernels is
dropping when the size of the window is increasing was to be
expected since in the limit of full sky coverage the $E$-$B$ mixing
disappears and the mixing kernel must go to zero.\\

\begin{figure}
\begin{center}
\leavevmode
\psfig {file=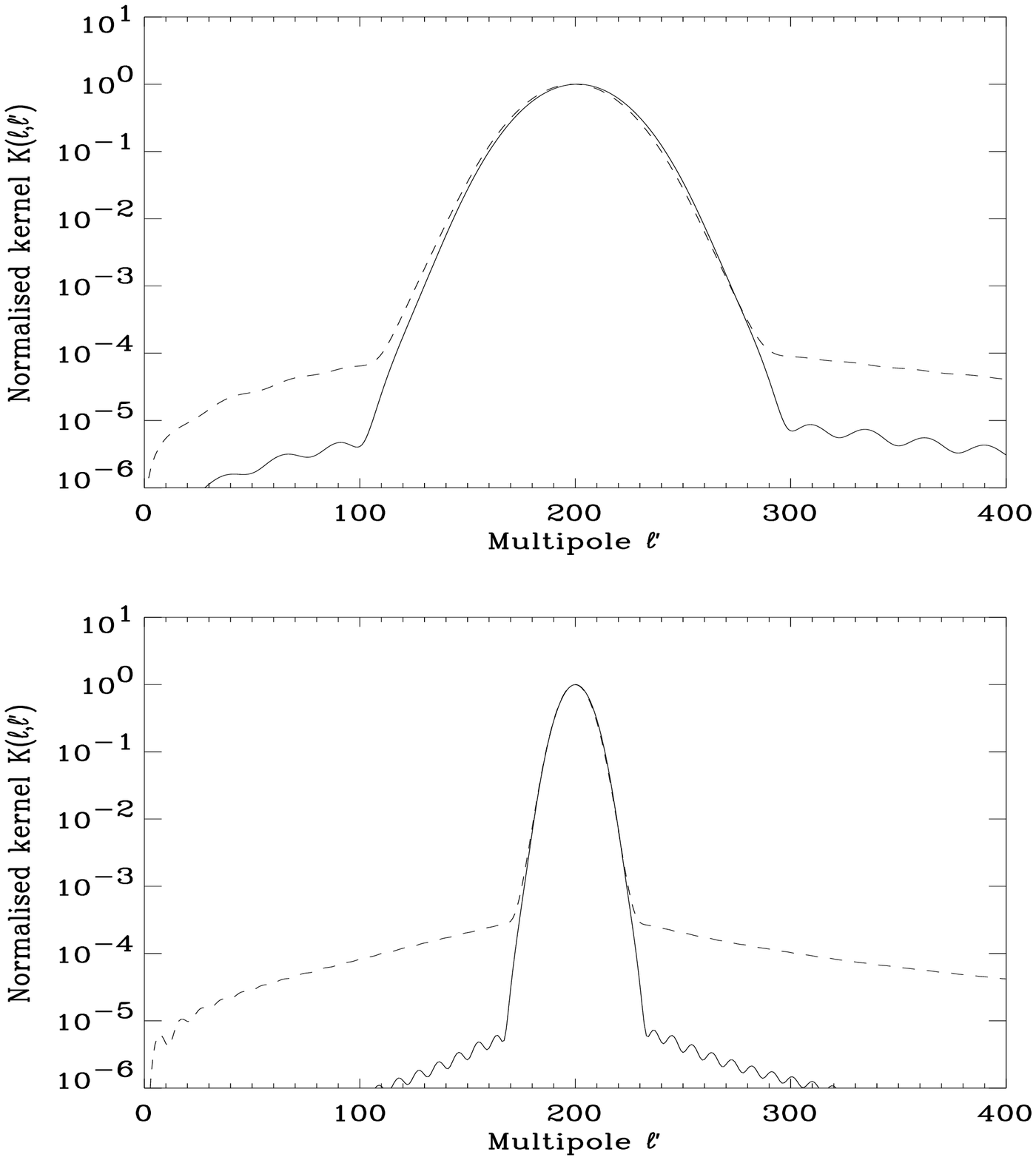,bbllx=0pt,bblly=150pt,bburx=596pt,bbury=842pt,height=12cm,width=14cm}
\caption{A slice at \protect{$\ell=200$} of the kernel
\protect{$K(\ell,\ell')$} connecting
the full sky temperature power spectrum with the cut sky temperature
power spectrum (solid line) and the kernel
\protect{$K_{-2}(\ell,\ell')$} which is mixing the $E$ and $B$ mode
polarisation power spectra on the cut sky. The upper plot
is for a $5$ degree FWHM Gaussian Gabor window and the lower plot for
a $15$ degree Gaussian window. The kernels are here normalised to $1$
at the peak at \protect{$\ell=200$} in order to compare the shapes of
the kernels.}
\label{fig:km2cutg}
\end{center}
\end{figure}

\begin{figure}
\begin{center}
\leavevmode
\psfig {file=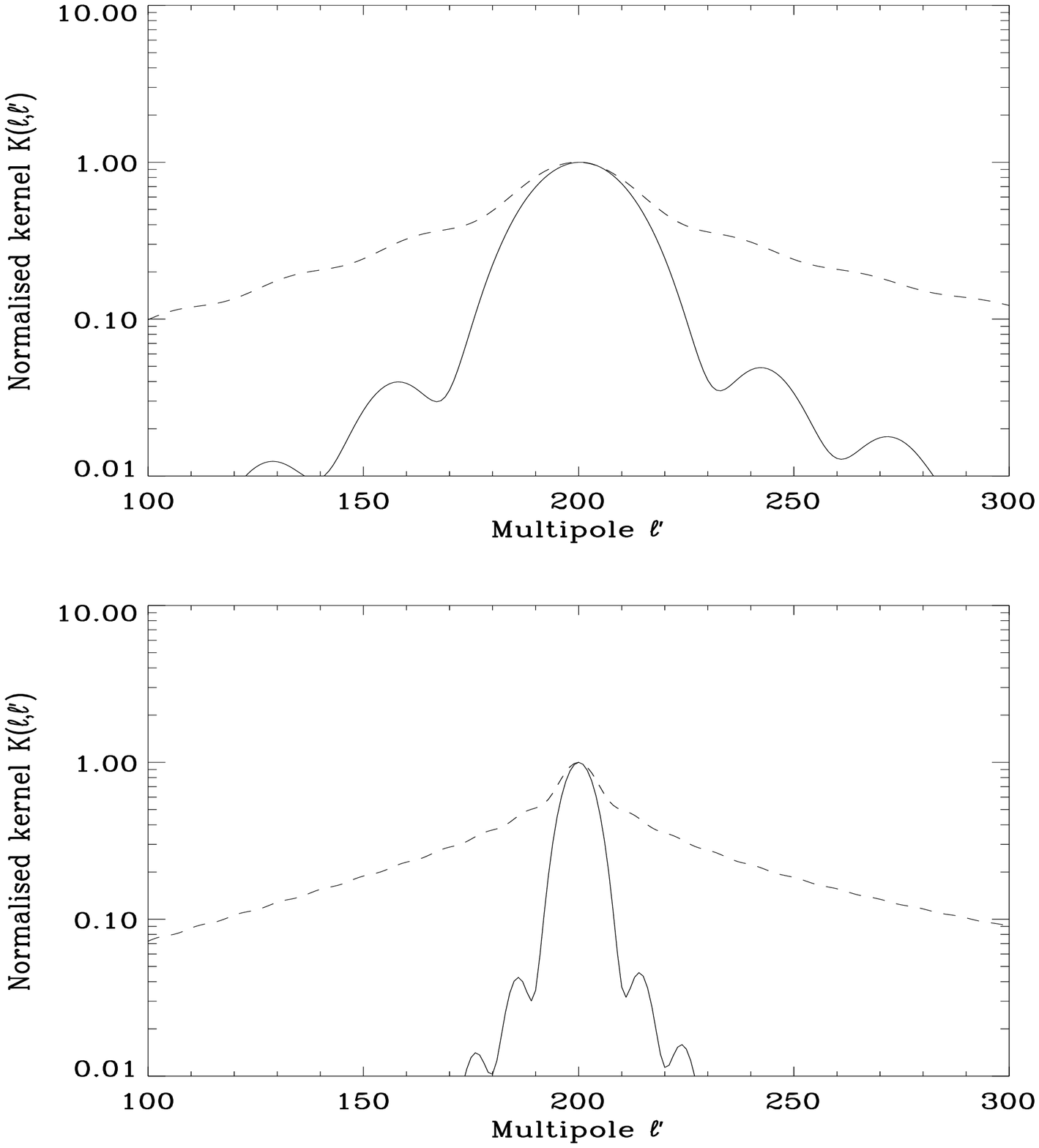,bbllx=0pt,bblly=150pt,bburx=596pt,bbury=842pt,height=12cm,width=14cm}
\caption{This figure is the same as figure (\ref{fig:km2cutg}) for
tophat windows covering the same area on the sky.}
\label{fig:km2cutth}
\end{center}
\end{figure}

In figure (\ref{fig:km2cutg}) a slice at
$\ell=200$ of the temperature kernel and the mixing kernel is
shown for the $5$ and $15$ degree FWHM Gaussian Gabor window. The
kernels are normalised to one at the peak so that the shapes can be
compared. For the Gaussian window, the shapes of the kernels still
seem to be the same. But the kernels for the corresponding tophat
windows shown in figure (\ref{fig:km2cutth}) does not have a Gaussian
shape and differs significantly from the other kernels.\\

\begin{figure}
\begin{center}
\leavevmode
\psfig {file=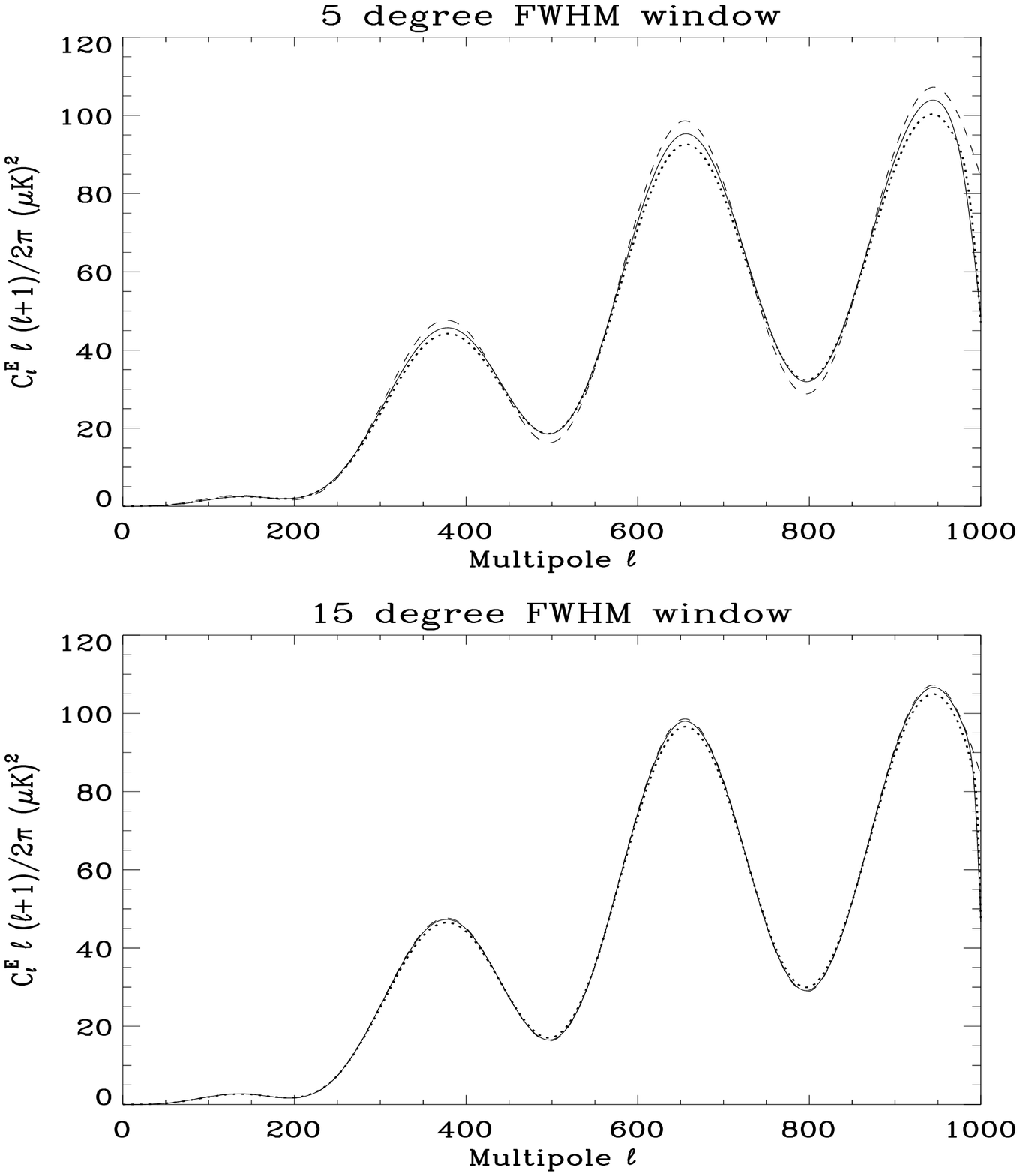,bbllx=0pt,bblly=150pt,bburx=696pt,bbury=842pt,height=12cm,width=16cm}
\caption{The  windowed polarisation power spectra \protect{$\tilde C_\ell^E$} for a $5$ and $15$ degree FWHM Gaussian
Gabor window cut at $\theta_C=3\sigma$ (solid line) and for a tophat window covering the same area
on the sky (dotted line). All spectra are normalised in such a way that they can be compared
directly with the full sky spectrum which is shown on each plot as a
dashed line. Only in the first plot are all three lines visible. In
the three last plots, the full sky spectrum and the Gaussian pseudo
spectrum (dashed and solid line) are hardly distinguishable.}
\label{fig:pcle}
\end{center}
\end{figure}

\begin{figure}\ \\
\begin{center}
\leavevmode
\psfig {file=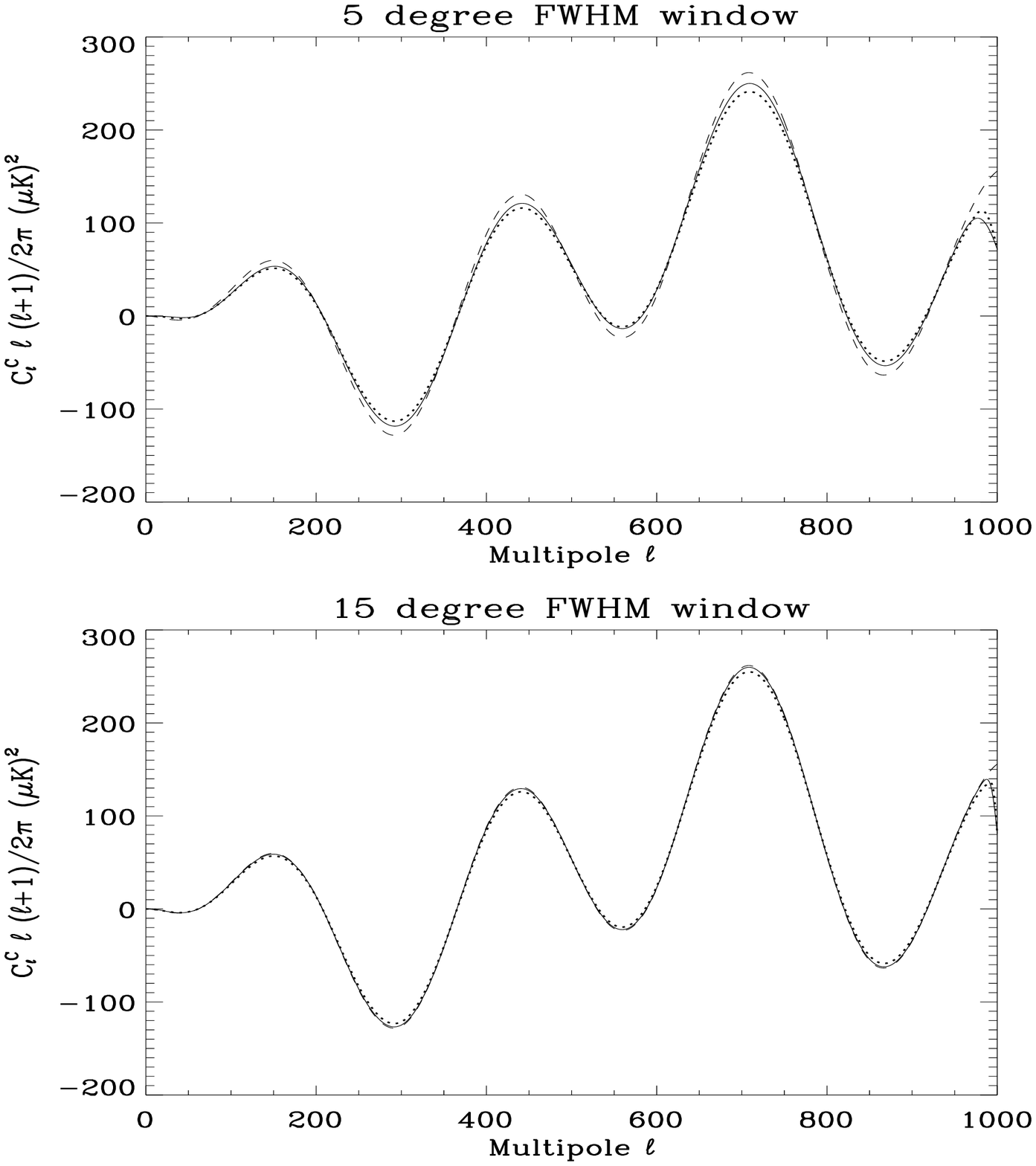,bbllx=0pt,bblly=150pt,bburx=696pt,bbury=842pt,height=12cm,width=16cm}
\caption{Same as figure (\ref{fig:pcle}) for the
temperature-polarisation cross power spectrum $C^C_\ell$.}
\label{fig:pclc}
\end{center}
\end{figure}

Since the kernels for the polarisation power spectra have a shape
similar to that of the temperature power spectrum the effect of a
Gabor window on the shape of the power spectrum should also be
similar. This can be seen in figure (\ref{fig:pcle}) and
(\ref{fig:pclc}). The figures show the full sky polarisation power
spectra (dashed line) $C_\ell^E$ (figure \ref{fig:pcle}) and $C_\ell^C$ (figure
\ref{fig:pclc}) for a standard CDM model. In this
model the $B$ component of polarisation is zero. On top of the full
sky power spectra we have plotted the polarisation pseudo power spectra
for a $5$ and $15$ degree Gaussian Gabor window (upper and lower plots
respectively) normalised so that it can be compared to the full sky
spectrum. The pseudo spectra for the corresponding tophat windows are
plotted as dotted lines. As expected the shape of the polarisation pseudo spectra
relative to the full sky spectra is similar to that for the temperature
spectrum shown in HGH. One difference is that the
polarisation pseudo spectra for the Gaussian window do not have the
characteristic extra peak at low multipole which is seen in the
temperature pseudo spectrum. This peak in the temperature spectrum
arose due to the steep $1/\ell(\ell+1)$ fall-off of the temperature
spectrum at low multipole. The polarisation spectra do not have this
steep
fall-off and for this reason there is no extra peak.\\

\begin{figure}
\begin{center}
\leavevmode
\psfig {file=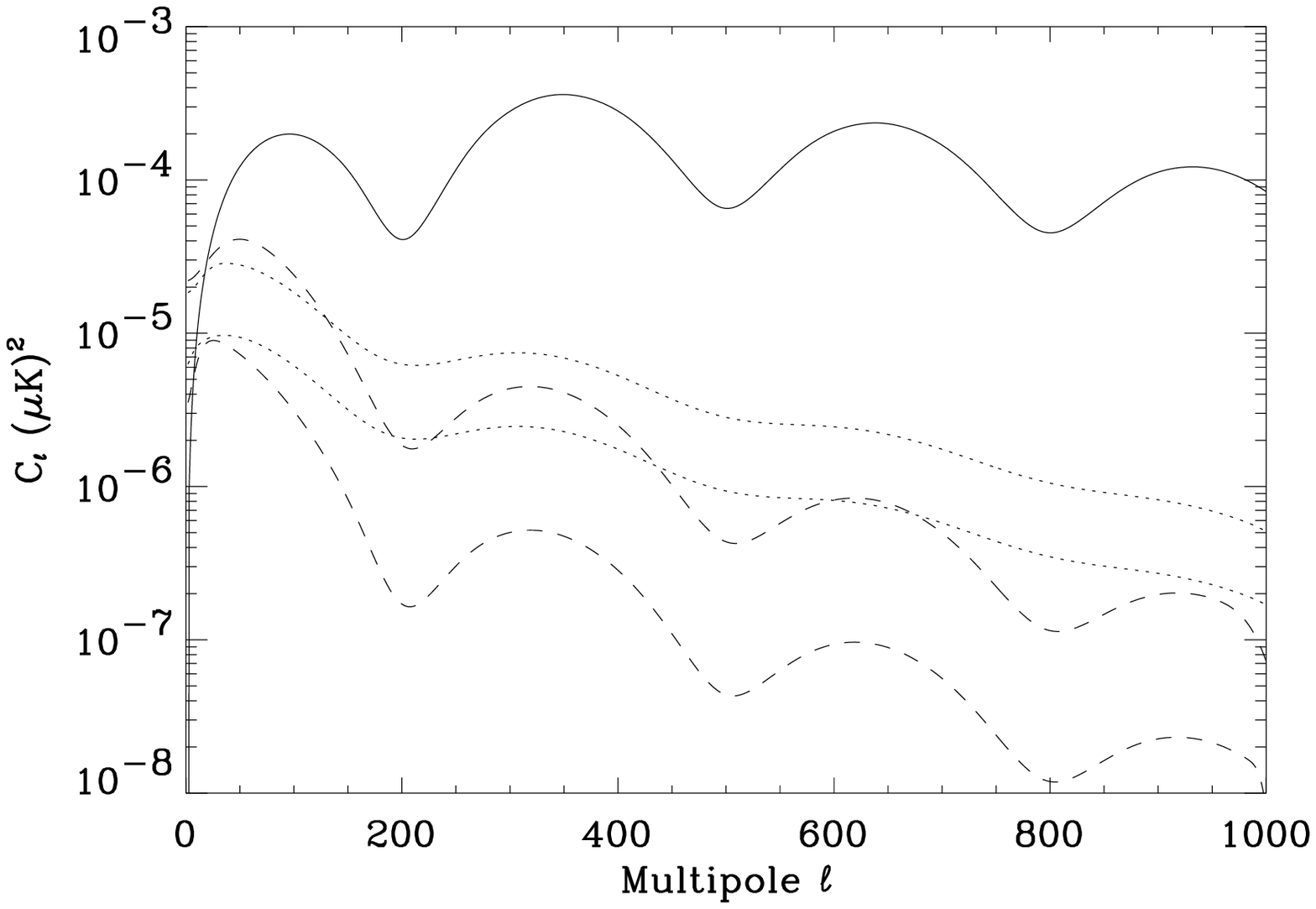,height=10cm,width=14cm}
\caption{The full sky $C^E_\ell$ power spectrum plotted together with
the spectra $\tilde C^B_\ell$ on the windowed sky. The dashed lines
show the $B$ spectra for a $5$ and $15$ degree FWHM Gaussian Gabor window
(upper and lower line respectively). The dotted lines are for the
corresponding tophat windows. The pseudo spectra are normalised so
that they can be compared directly with the full sky spectrum. In the
model used, there was no $B$ polarisation spectrum for the full
sky. The $\tilde C^B_\ell$ shown arise due to the mixing of $E$ and
$B$ modes on the cut sky only.}
\label{fig:pclb}
\end{center}
\end{figure}

Because of the mixing of $E$ and $B$ modes there is also a $B$ polarisation component
$\tilde C^B_\ell$ for the pseudo spectrum even when the input full sky
$C^B_\ell$ were zero. This is shown in figure (\ref{fig:pclb}) where we
have plotted the full sky spectrum $C^E_\ell$ and the pseudo spectra
$\tilde C_\ell^B$ for the $5$ and $15$ degree FWHM Gaussian Gabor windows
and corresponding tophat windows. The pseudo spectra are normalised
so that they can be compared directly to the full sky spectrum. The dashed lines show the pseudo
spectra for the Gaussian window. The upper line is for the $5$ degree
window and the lower line for the $15$ degree window. As expected the
size of the $B$ component is dropping with increasing window size. The
$\tilde C^B_\ell$ for the tophat windows are plotted as dotted
lines. The shape of the pseudo spectra $\tilde C^B_\ell$ for the Gaussian windows are
roughly following the shape of the full sky $C^E_\ell$. This could be
expected because the mixing kernel $K_{-2}(\ell,\ell')$ for the
Gaussian window has a Gaussian shape close to the diagonal, similar to
the other kernels (see figure (\ref{fig:km2cutg})). The pseudo spectra
$C^B_\ell$ for the tophat windows however are much smoother due to the
much broader kernel (figure \ref{fig:km2cutth}).

In the same way as for the temperature power spectrum, we have shown that
the polarisation pseudo spectra resemble the full sky polarisation
spectra when the patches on the sky are large enough. This motivates
the use of the polarisation pseudo power spectra as input to a
likelihood estimation of the polarisation power spectra in the same
way as for the temperature power spectrum showed in HGH. To do likelihood analysis one needs to
find theoretical expressions for the correlations between different
$\tilde C_\ell^Z$ (Z=\{T,E,C\}).

\section{Likelihood Analysis}
\label{sect:likformpol}

In HGH a Gaussian likelihood ansatz with the pseudo power spectrum as the input data was successfully used to estimate the power spectrum. Because of the similarities
between the kernels of the polarisation power spectra and the
temperature power spectrum we can assume that this works for the estimation of the polarisation power spectra as well. We will now show the results of some Monte Carlo
simulations confirming this assumption. In this section we will assume that the $B$
component of polarisation is so small that it can be neglected. We will
only concentrate on the $T$, $E$ and $C$ components as in most standard theories of the early universe the $B$ component will be too small to be measured by the MAP and Planck satellite experiments \cite{jaffe0}.\\

\begin{figure}
\begin{center}
\leavevmode
\psfig {file=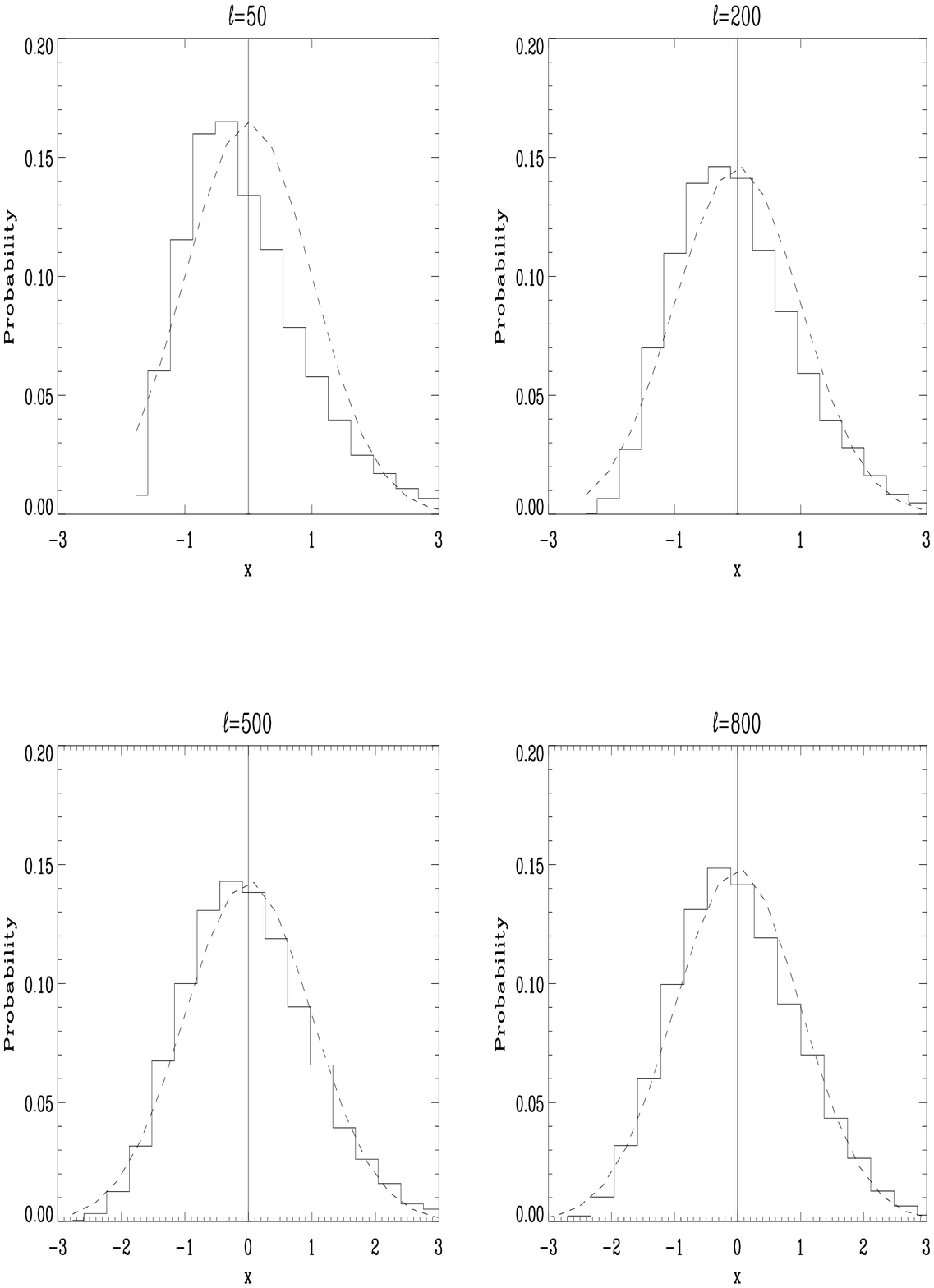,bbllx=0pt,bblly=0pt,bburx=650pt,bbury=842pt,height=12cm,width=16cm}
\caption{The probability distribution of \protect{$\tilde C^E_\ell$}
taken from 10000 simulations with a \protect{$5^\circ$} FWHM Gaussian
Gabor window truncated at $\theta_C=3\sigma$. The variable $x$ is given as
\protect{$x=(\tilde C^E_\ell-\VEV{\tilde C^E_\ell})/\sqrt{\VEV{(\tilde C^E_\ell-\VEV{\tilde
C^E_\ell})^2}}$}. The dashed line is a Gaussian
with the theoretical mean and standard deviation of the
\protect{$\tilde C^E_\ell$}. The plot shows the \protect{$\tilde C^E_\ell$}
distribution for \protect{$\ell=50$, $\ell=200$, $\ell=500$},
and \protect{$\ell=800$}. The probabilities are normalised such that
the integral over $x$ is $1$.}
\label{fig:probpcleg5}
\end{center}
\end{figure}

\begin{figure}
\begin{center}
\leavevmode
\psfig {file=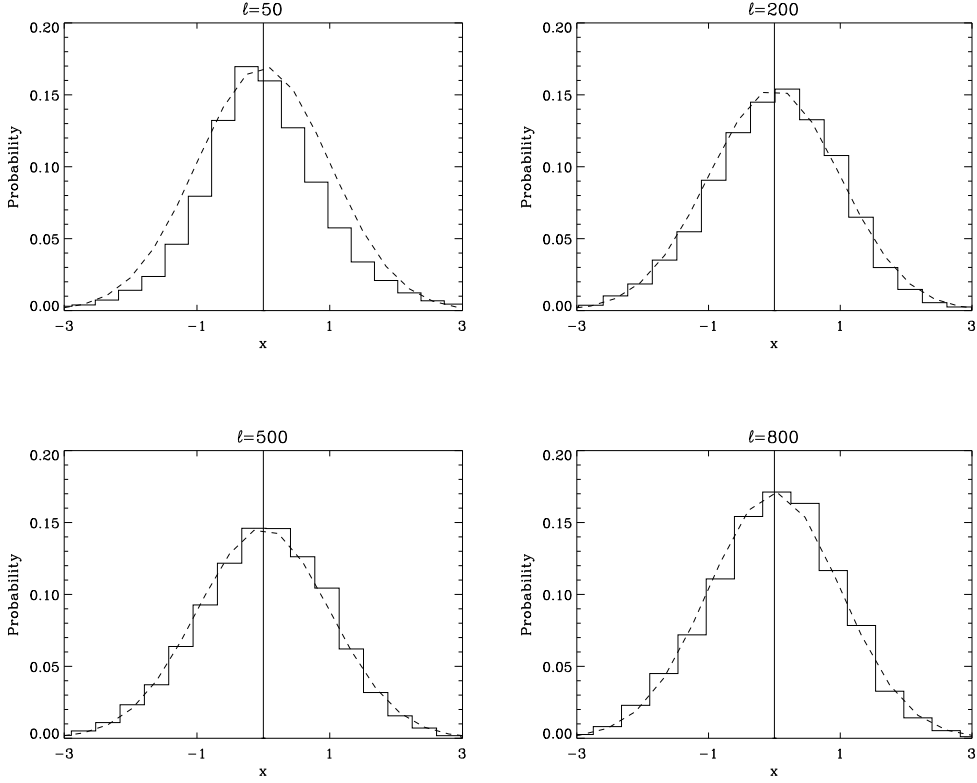,bbllx=0pt,bblly=0pt,bburx=650pt,bbury=842pt,height=12cm,width=16cm}
\caption{Same as figure (\ref{fig:probpcleg5}) for the
temperature-polarisation cross spectra $\tilde C^C_\ell$.}
\label{fig:probpclcg5}
\end{center}
\end{figure}

\begin{figure}
\begin{center}
\leavevmode
\psfig {file=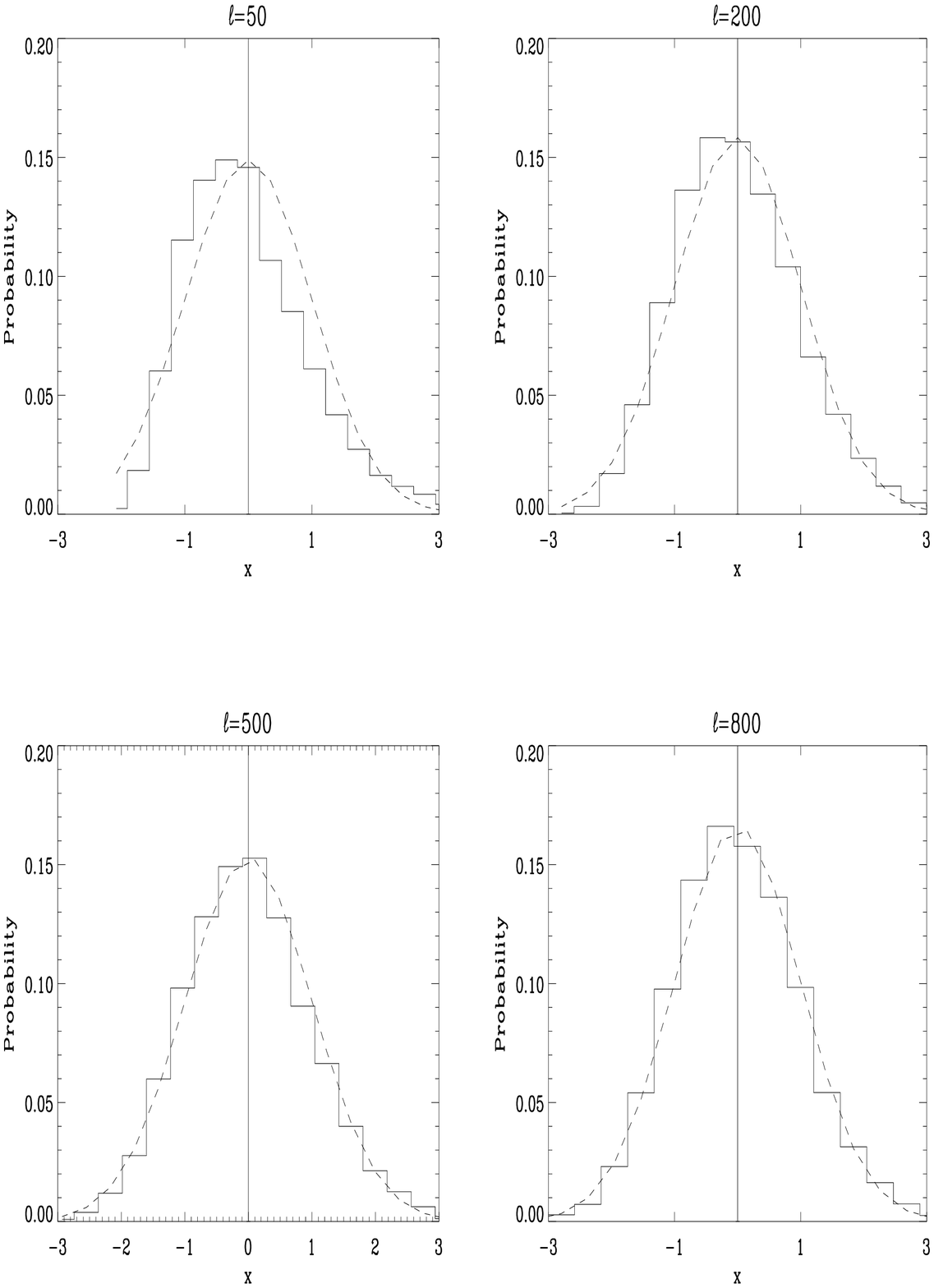,bbllx=0pt,bblly=0pt,bburx=650pt,bbury=842pt,height=12cm,width=16cm}
\caption{Same as figure (\ref{fig:probpcleg5}) for a $15^\circ$
FWHM Gaussian Gabor window.}
\label{fig:probpcleg15}
\end{center}
\end{figure}

\begin{figure}
\begin{center}
\leavevmode
\psfig {file=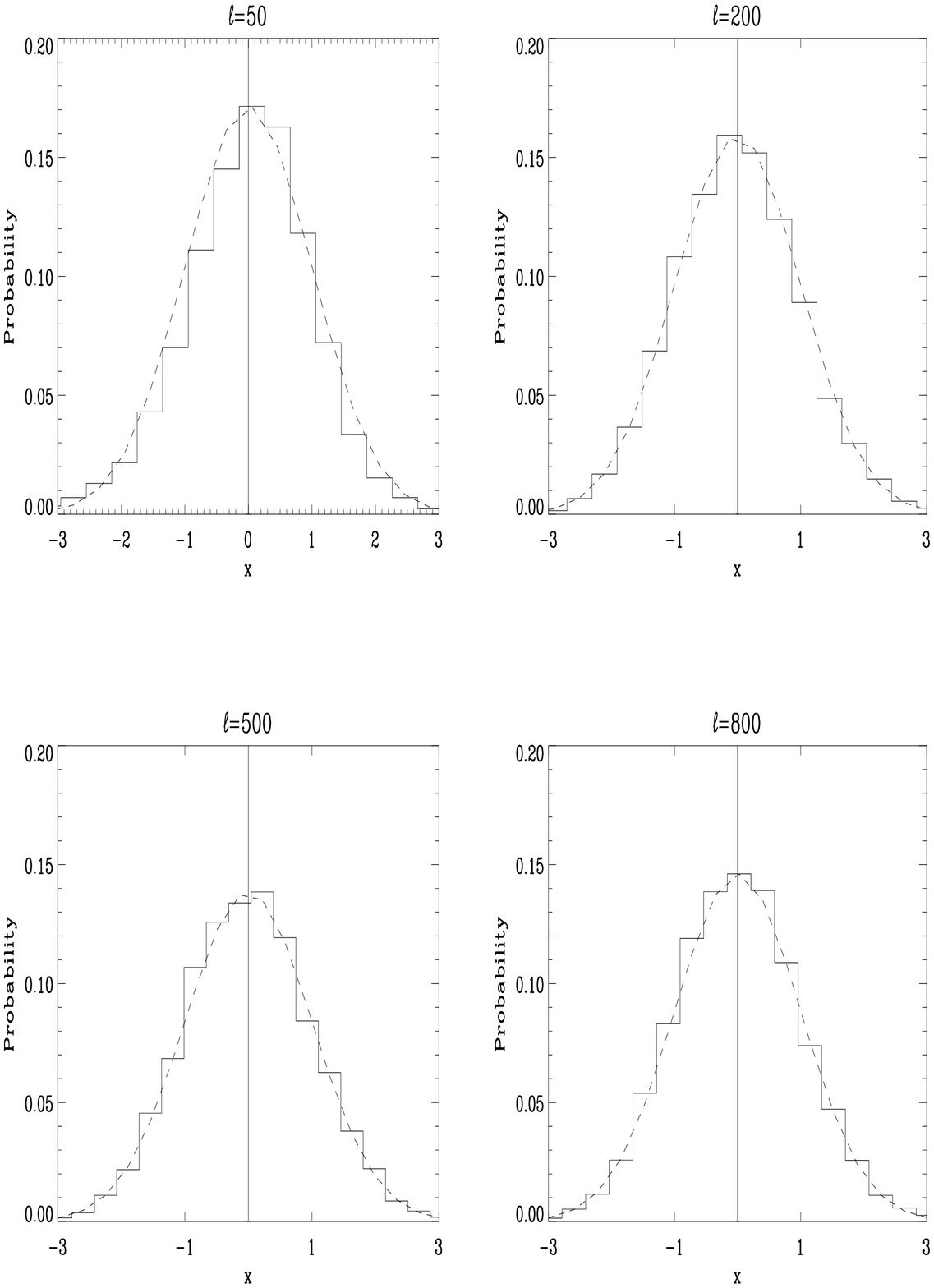,bbllx=0pt,bblly=0pt,bburx=650pt,bbury=842pt,height=12cm,width=16cm}
\caption{Same as figure (\ref{fig:probpclcg5}) for a $15^\circ$
FWHM Gaussian Gabor window.}
\label{fig:probpclcg15}
\end{center}
\end{figure}

In figure (\ref{fig:probpcleg5}) and (\ref{fig:probpclcg5}) we have
plotted the probability distribution of the $\tilde C^E_\ell$ and
$\tilde C^C_\ell$ from 10000 simulations. The probability distribution
(histogram) is plotted on top of a Gaussian (dashed line) with mean
and FWHM taken from the theoretical expressions derived in Appendix (\ref{app:polcormat}). In these simulations we were
using a $5$ degree FWHM Gaussian Gabor window with $\theta_C=3\sigma$. In
figure (\ref{fig:probpcleg15}) and (\ref{fig:probpclcg15}) we show the
results of similar simulations with a $15$ degree FWHM Gaussian Gabor
window. As expected the trend is that the distributions get more and
more Gaussian for higher multipoles and for bigger windows. For the
$15^\circ$ FWHM window, the distribution is very close to a Gaussian
for the multipoles above $\ell=50$.

\begin{figure}
\begin{center}
\leavevmode
\psfig {file=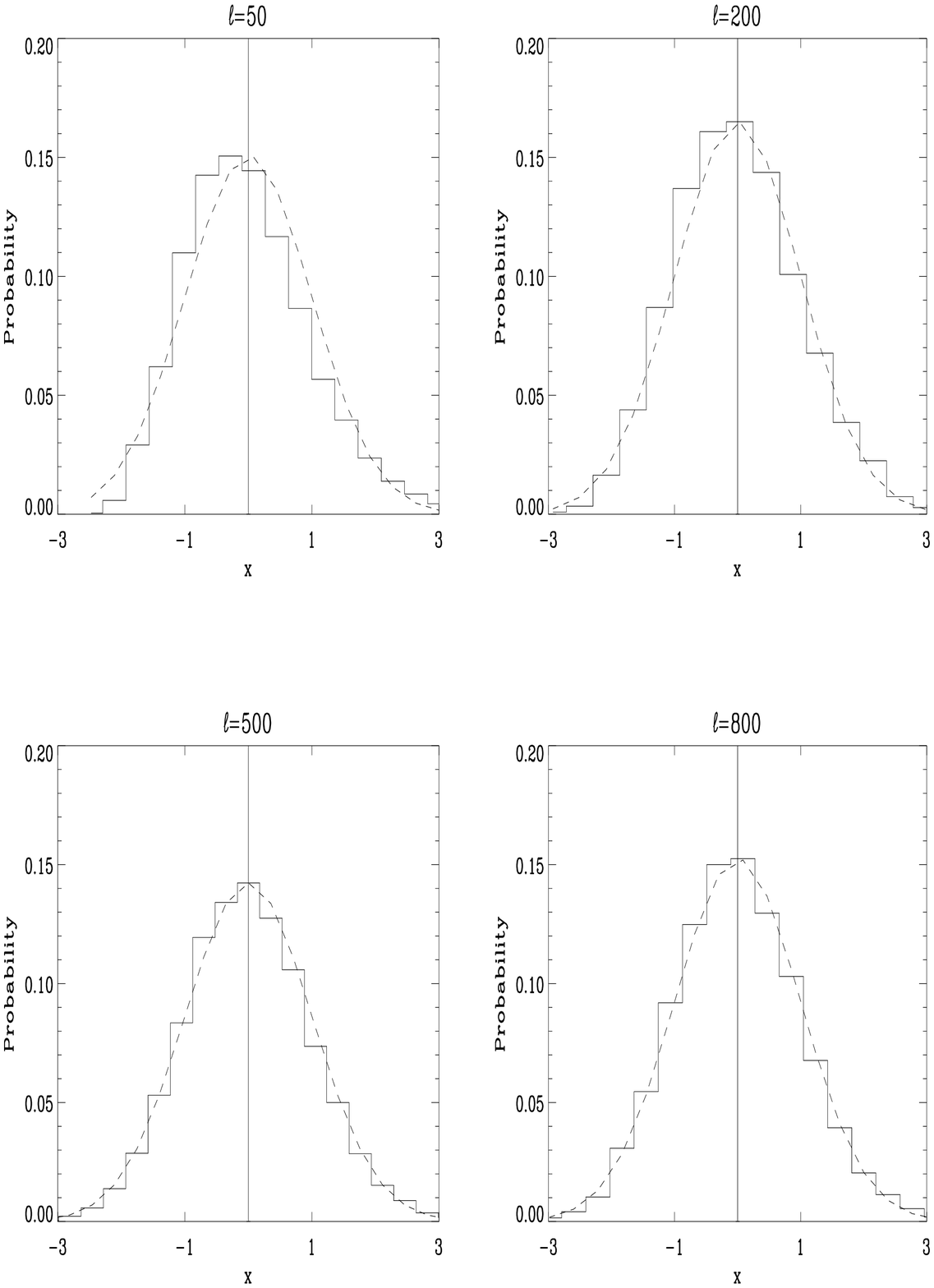,bbllx=0pt,bblly=0pt,bburx=650pt,bbury=842pt,height=12cm,width=16cm}
\caption{Same as figure (\ref{fig:probpcleg15}) for a tophat window
covering the same area on the sky}
\label{fig:probpcleth15}
\end{center}
\end{figure}

\begin{figure}
\begin{center}
\leavevmode
\psfig {file=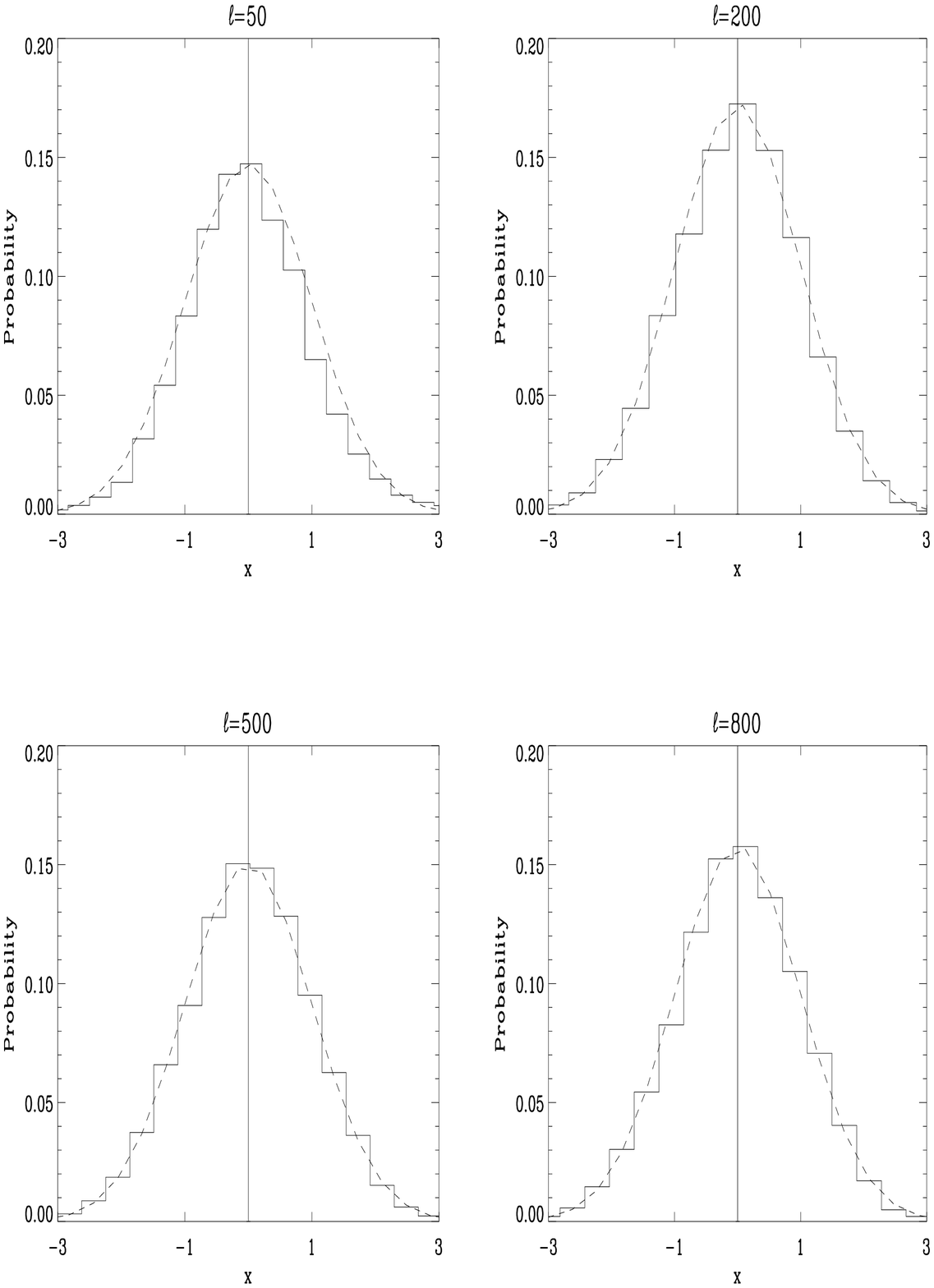,bbllx=0pt,bblly=0pt,bburx=650pt,bbury=842pt,height=12cm,width=16cm}
\caption{Same as figure (\ref{fig:probpclcg15}) for a tophat window
covering the same area on the sky}
\label{fig:probpclcth15}
\end{center}
\end{figure}

Figure (\ref{fig:probpcleth15}) and (\ref{fig:probpclcth15}) show the
probability distribution for a tophat window covering the same area on
the sky as the Gaussian window used in figure (\ref{fig:probpcleg15})
and (\ref{fig:probpclcg15}). Also this distribution is very close to a Gaussian.\\

The previous plots have shown that a Gaussian likelihood ansatz for
the polarisation pseudo spectra seems to be a very good approximation
provided that the window is big enough. As for the temperature
spectrum, the approximation is no longer valid for the lowest
multipoles, but as was shown for the temperature power spectrum, this
might only give rise to a very small downward bias for the estimates
of the lowest multipoles.\\

\begin{figure}
\begin{center}
\leavevmode
\psfig {file=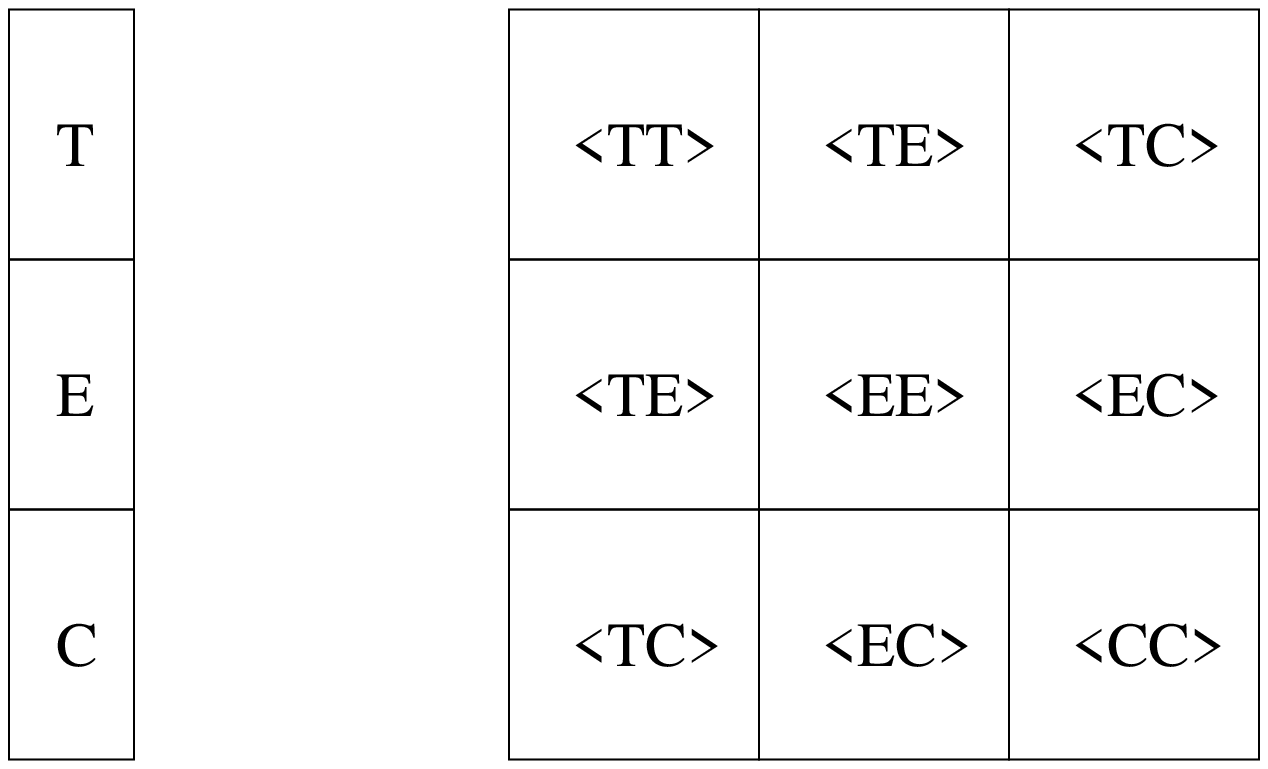,bbllx=200pt,bblly=250pt,bburx=472pt,bbury=592pt,height=10cm,width=10cm,angle=270}
\caption{The figure shows the structure of the datavector $\mathbf{d}$ on
the left hand side and the correlation matrix $\mt{M}$ on the right
hand side used for joint likelihood estimation of temperature and
polarisation power spectra.} 
\label{fig:polfig}
\end{center}
\end{figure}

The form of the log-likelihood to minimise is therefore still
\begin{equation}
L=\mathbf{d}^T \mt{M}^{-1}\mathbf{d}+\ln{\det{\mt{M}}},
\end{equation}
where the data vector now consists of the temperature
and polarisation power spectra $\mathbf{d}=\{\mathbf{d}^T,\mathbf{d}^E,\mathbf{
d}^C\}$. Here the $\mathbf{d}^Z$ vectors are given as
\begin{equation}
d_i^Z=\tilde C_{\ell_i}^Z-\VEV{\tilde C_{\ell_i}^Z},
\end{equation}
where $Z=\{T,E,B\}$. Similarly the correlation matrix $\mt{M}$ will
consist of blocks $\mt{M}_{ZZ'}$ defined as
\begin{equation}
M_{ZZ',ij}=\VEV{\tilde C_{\ell_i}^Z\tilde C_{\ell_j}^{Z'}}-\VEV{\tilde
C_{\ell_i}^Z}\VEV{\tilde C_{\ell_j}^{Z'}}.
\end{equation}
This structure of the data vector and correlation matrix is shown in
figure (\ref{fig:polfig}).\\

For fast likelihood estimation, it is crucial that one can calculate
the average pseudo spectra $\VEV{\tilde C^Z_\ell}$ and correlation matrix
$\mt{M}$ fast. The formalism in HGH which enabled
fast calculations of these quantities for the temperature power
spectrum is extended to polarisation in Appendix (\ref{app:polcormat}) (signal) and Appendix (\ref{app:polnoise}) (noise).

\begin{figure}
\begin{center}
\leavevmode
\psfig {file=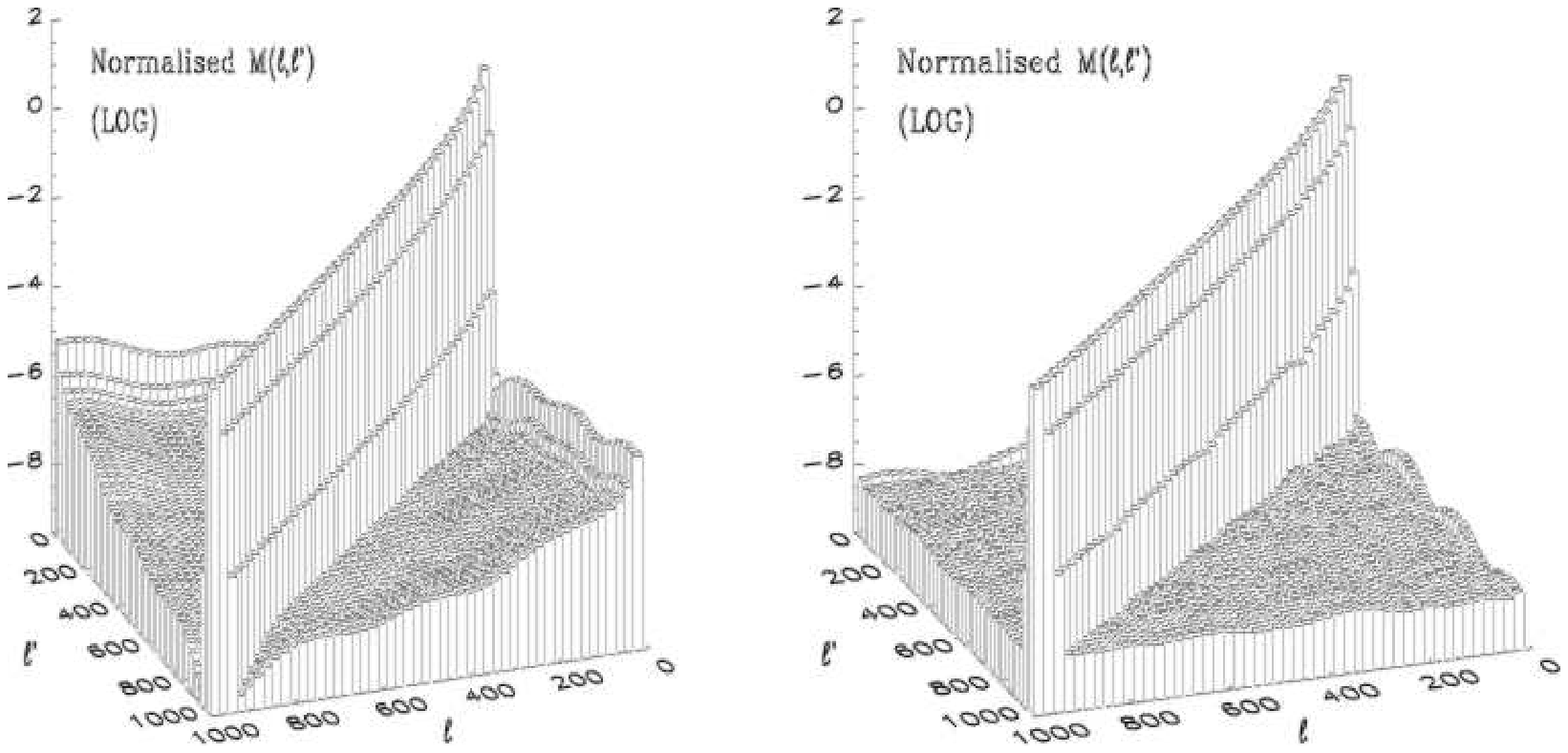,height=10cm,width=14cm}
\caption{The correlation matrices $M(\ell,\ell')$ in the figure show the correlations between the
temperature pseudo power spectrum coefficients and between the $E$
mode polarisation pseudo spectrum coefficients for a $15$ degree FWHM
Gaussian Gabor window. The left plot
shows \protect{$(\VEV{\tilde C^T_\ell\tilde C^T_{\ell'}}-\VEV{\tilde C^T_\ell}\VEV{\tilde
C^T_{\ell'}})/(\VEV{\tilde C^T_\ell}\VEV{\tilde C^T_{\ell'}})$} and the right
plot shows \protect{$(\VEV{\tilde C^E_\ell\tilde C^E_{\ell'}}-\VEV{\tilde C^E_\ell}\VEV{\tilde
C^E_{\ell'}})/(\VEV{\tilde C^E_\ell}\VEV{\tilde C^E_{\ell'}})$}. A standard
CDM power spectrum was used to produce the plots.} 
\label{fig:corttee}
\end{center}
\end{figure}

In figure (\ref{fig:corttee}) we have plotted the signal correlation matrix 
$M^{TT}_{\ell\ell'}$ next to the matrix $M^{EE}_{\ell\ell'}$. A
standard CDM power spectrum without $B$ mode polarisation was used. The two matrices
are very similar. One big difference is that the matrix for $E$ mode
polarisation is missing the 'wall' at low multipoles present in the
temperature matrix. As discussed before this is because of the
different shapes for the $T$ and $E$ power spectra at low
multipoles. The temperature power spectrum drops steeply at low
$\ell$ while this is not the case for the $E$ mode polarisation spectrum.\\

\begin{figure}
\begin{center}
\leavevmode
\psfig {file=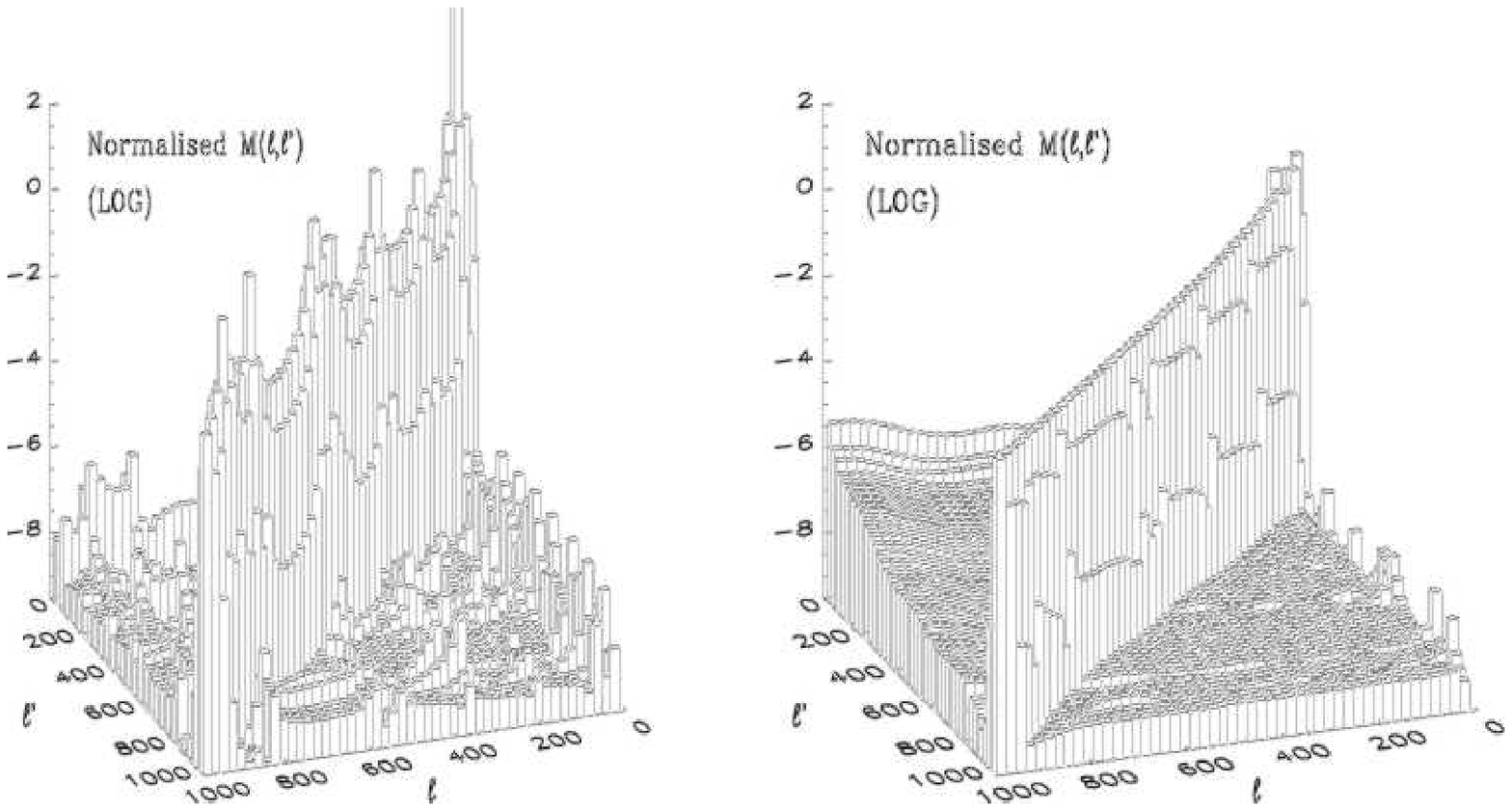,height=10cm,width=14cm}
\caption{The correlation matrices $M(\ell,\ell')$ in the figure show the correlations between the
cross-correlation pseudo power spectrum $C$ coefficients and between
the temperature and
cross correlation pseudo spectrum coefficients $C$ for a $15$ degree FWHM
Gaussian Gabor window. The left plot
shows \protect{$(\VEV{\tilde C^C_\ell\tilde C^C_{\ell'}}-\VEV{\tilde C^C_\ell}\VEV{\tilde
C^C_{\ell'}})/(\VEV{\tilde C^C_\ell}\VEV{\tilde C^C_{\ell'}})$} and the right
plot shows \protect{$(\VEV{\tilde C^T_\ell\tilde C^C_{\ell'}}-\VEV{\tilde C^T_\ell}\VEV{\tilde
C^C_{\ell'}})/(\VEV{\tilde C^T_\ell}\VEV{\tilde C^C_{\ell'}})$}. A standard
CDM power spectrum was used to produce the plots.} 
\label{fig:corcctc}
\end{center}
\end{figure}

\begin{figure}
\begin{center}
\leavevmode
\psfig {file=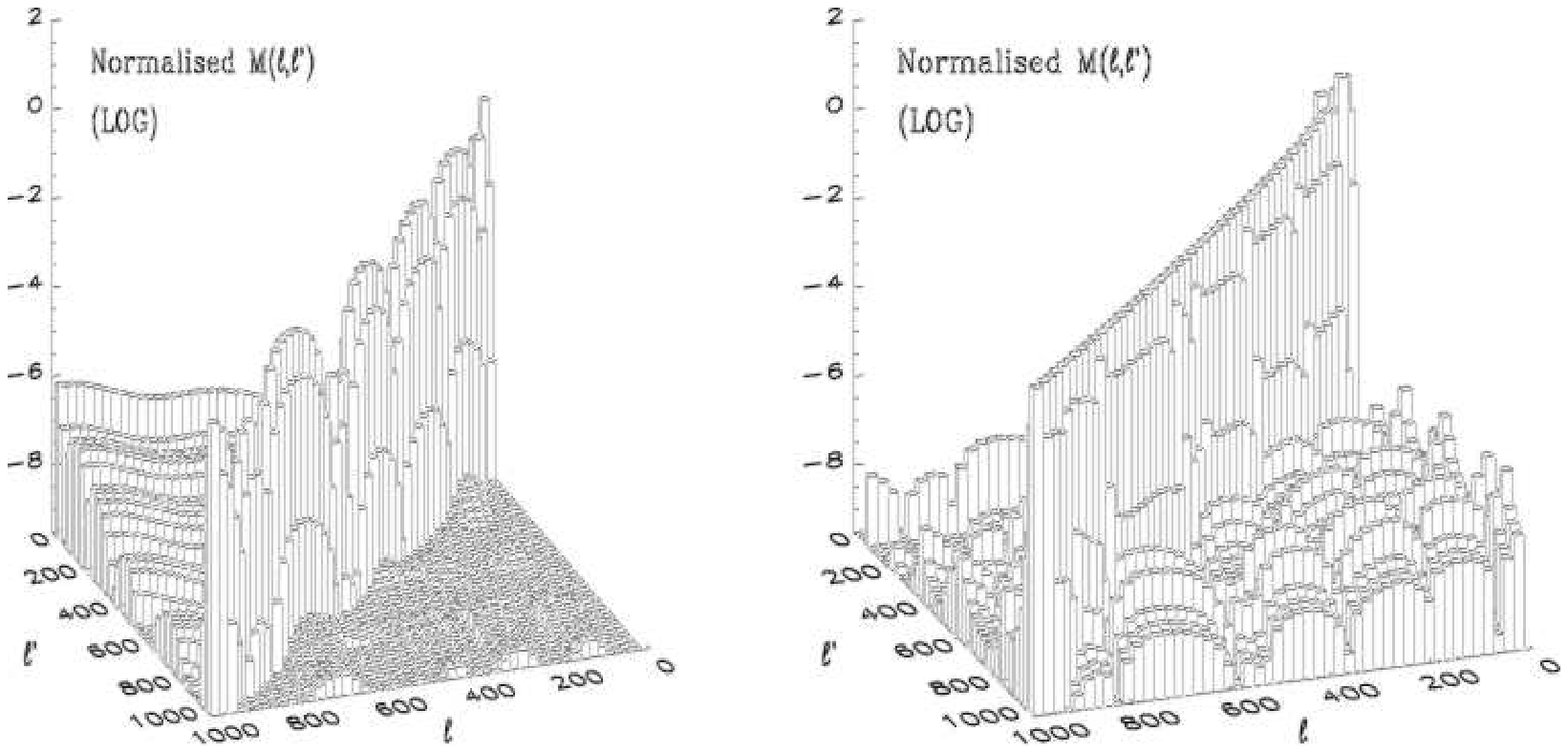,height=10cm,width=14cm}
\caption{The correlation matrices $M(\ell,\ell')$ in the figure show the correlations between the
temperature and $E$ mode polarisation pseudo spectrum coefficients and between
the $E$ mode polarisation and
cross correlation pseudo spectrum coefficients $C$ for a $15$ degree FWHM
Gaussian Gabor window. The left plot
shows \protect{$(\VEV{\tilde C^T_\ell\tilde C^E_{\ell'}}-\VEV{\tilde C^T_\ell}\VEV{\tilde
C^E_{\ell'}})/(\VEV{\tilde C^T_\ell}\VEV{\tilde C^E_{\ell'}})$} and the right
plot shows \protect{$(\VEV{\tilde C^E_\ell\tilde C^C_{\ell'}}-\VEV{\tilde C^E_\ell}\VEV{\tilde
C^C_{\ell'}})/(\VEV{\tilde C^E_\ell}\VEV{\tilde C^C_{\ell'}}
)$}. A standard
CDM power spectrum was used to produce the plots.} 
\label{fig:corteec}
\end{center}
\end{figure}

In figures (\ref{fig:corcctc}) and (\ref{fig:corteec}) the
$M^{CC}_{\ell\ell'}$, $M^{TC}_{\ell\ell'}$, $M^{TE}_{\ell\ell'}$ and
$M^{EC}_{\ell\ell'}$ matrices are shown. All matrices are diagonally
dominant and since the values on the diagonals all have the same order of magnitude, they all have to be included in the total matrix $M$.

\section{Results of Likelihood Estimations}

The likelihood estimation was carried out in the same way as for the
temperature power spectrum. As discussed in HGH, when we observe the cut-sky we do not have enough information to estimate the full-sky $C_\ell$ for all multipoles. One has to estimate the $C_\ell$ in $N^{bin}$ bins. Also, the $\tilde C_\ell$ coefficients are not independent on the cut sky, so a limited number $N^{in}\leq N^{bin}$ of pseudo spectrum coefficients have to be used as input to the likelihood. How many multipoles depends on the width $\Delta\ell$ of the Gabor kernel for a given window which is discussed in more detail in HGH.\\

The power spectra were estimated in bins
defined as
\begin{eqnarray}
C^T_\ell&=&\frac{D^T_b}{\ell(\ell+1)},\ \
\ell_b\leq\ell<\ell_{b+1},\\
C^E_\ell&=&\frac{D^E_b}{\ell(\ell+1)},\ \
\ell_b\leq\ell<\ell_{b+1},\\
\end{eqnarray}
where $\ell_b$ is the first multipole in bin $b$.
A similar binning does not work for for the temperature-polarisation
cross correlation power spectrum. The reason for this is the Schwarz
inequality $C^C_\ell\leq\sqrt{C^T_\ell C^E_\ell}$. During likelihood
maximisation one must make sure that the estimated value of $C^C_\ell$
never exceeds $\sqrt{C^T_\ell C^E_\ell}$. The way we solved this
problem was to estimate for $C^C_\ell/\sqrt{C^T_\ell C^E_\ell}$ under
the constraint that this value never exceeds $1$. So the binning is
then
\begin{eqnarray}
C^C_\ell&=&D^C_b\sqrt{C^T_\ell C^E_\ell}\\
&=&\frac{D^C_b\sqrt{D^T_bD^E_b}}{\ell(\ell+1)},
\end{eqnarray}
where as before $\ell_b\leq\ell<\ell_{b+1}$.\\

As an example we simulated a sky using $N_{side}=512$ resolution in
Healpix \cite{healpix} and a $10'$ beam. We added non-uniform noise to the map. A
reasonable assumption about the size of the noise deviations for
polarisation is to take $\sigma^P_j=\sqrt{2}\sigma^T_j$
\cite{pol1}. This is what we used in this test, but note that the formalism do not require any relation between $\sigma_T$ and $\sigma_P$. The noise level was set so that the
signal to noise ration for the temperature power spectrum was always
well above $1$ below the maximum multipole $\ell=1024$ whereas for the $E$ mode polarisation power spectrum it
was mostly below $1$ (see figure (\ref{fig:polpclnoavg})). This is close to
the values expected for the Planck HFI $143 GHz$ \cite{plancka} channel. For the
analysis we used a $15$ degree Gaussian Gabor window. The result of one
single estimation is shown in figure (\ref{fig:polpclnoavg}). In this estimation we used $N^{in}=100$ pseudo spectrum multipoles as input and estimated for $N^{bin}=20$ multipole bins for each of the $T$, $E$ and $C$ modes.\\

\begin{figure}
\begin{center}
\leavevmode
\psfig {file=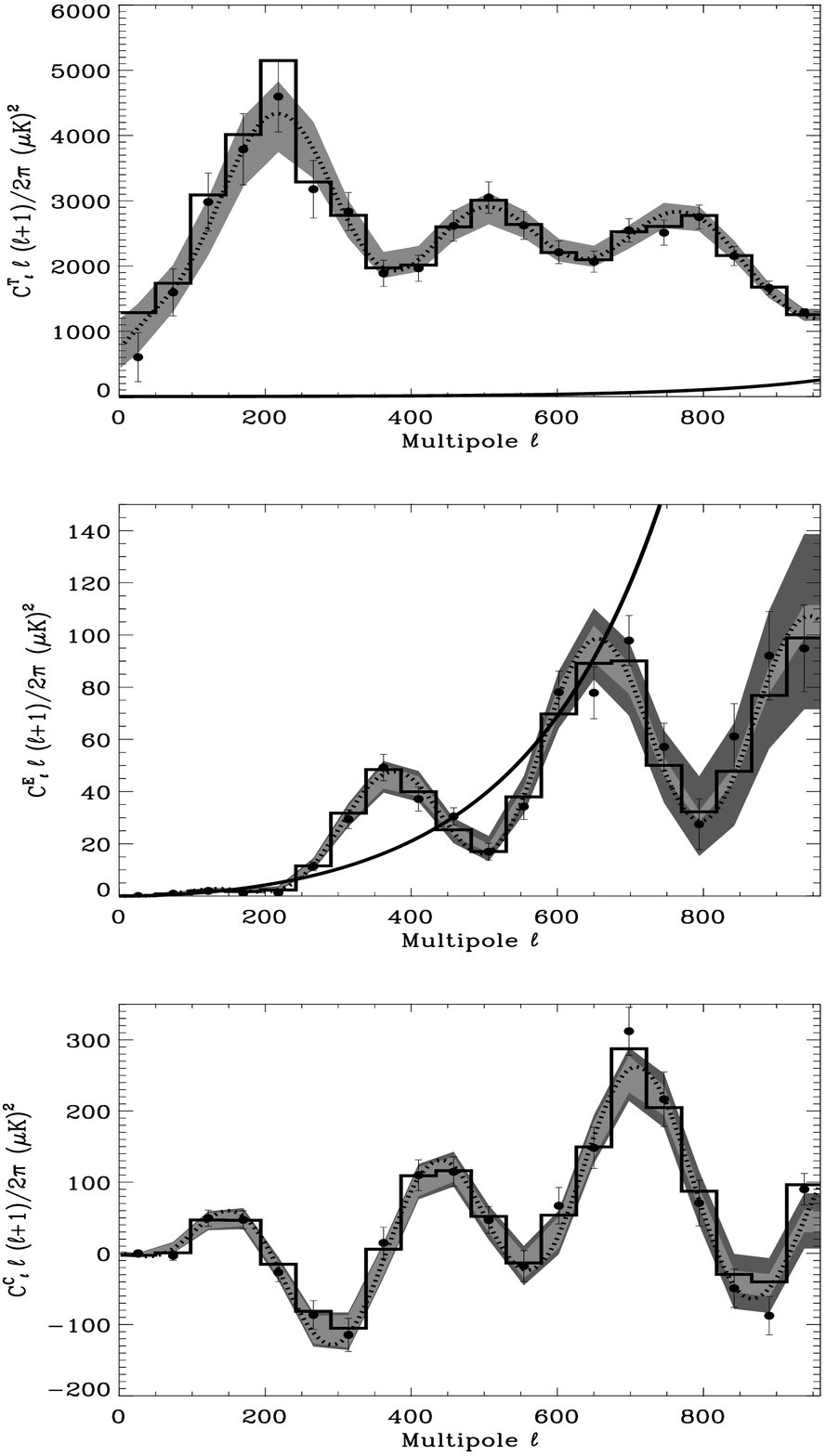,bbllx=0pt,bblly=0pt,bburx=696pt,bbury=842pt,height=18cm,width=22cm}
\caption{The result of a joint likelihood estimation of the temperature
power spectrum (upper plot) and the $E$ (middle plot) and $C$ (lower
plot) polarisation power spectra. The dotted line shows the full sky
average spectrum. The histogram shows the binned input
pseudo spectrum without noise. The shaded areas around the binned average full sky
power spectrum (not shown) show the expected deviations from
the average using the approximate formula for uniform noise. The bright
shaded area shows the cosmic and sample variance only whereas the dark
shaded area also shows expected variance due to noise. The dots show
the estimate with $1\sigma$ error bars taken from the inverse Fisher
matrix. In the analysis a $15$ degree FWHM Gaussian Gabor
window with a $\theta_C=3\sigma$ cutoff was
used.}
\label{fig:polpcl}
\end{center}
\end{figure}

\begin{figure}
\begin{center}
\leavevmode
\psfig {file=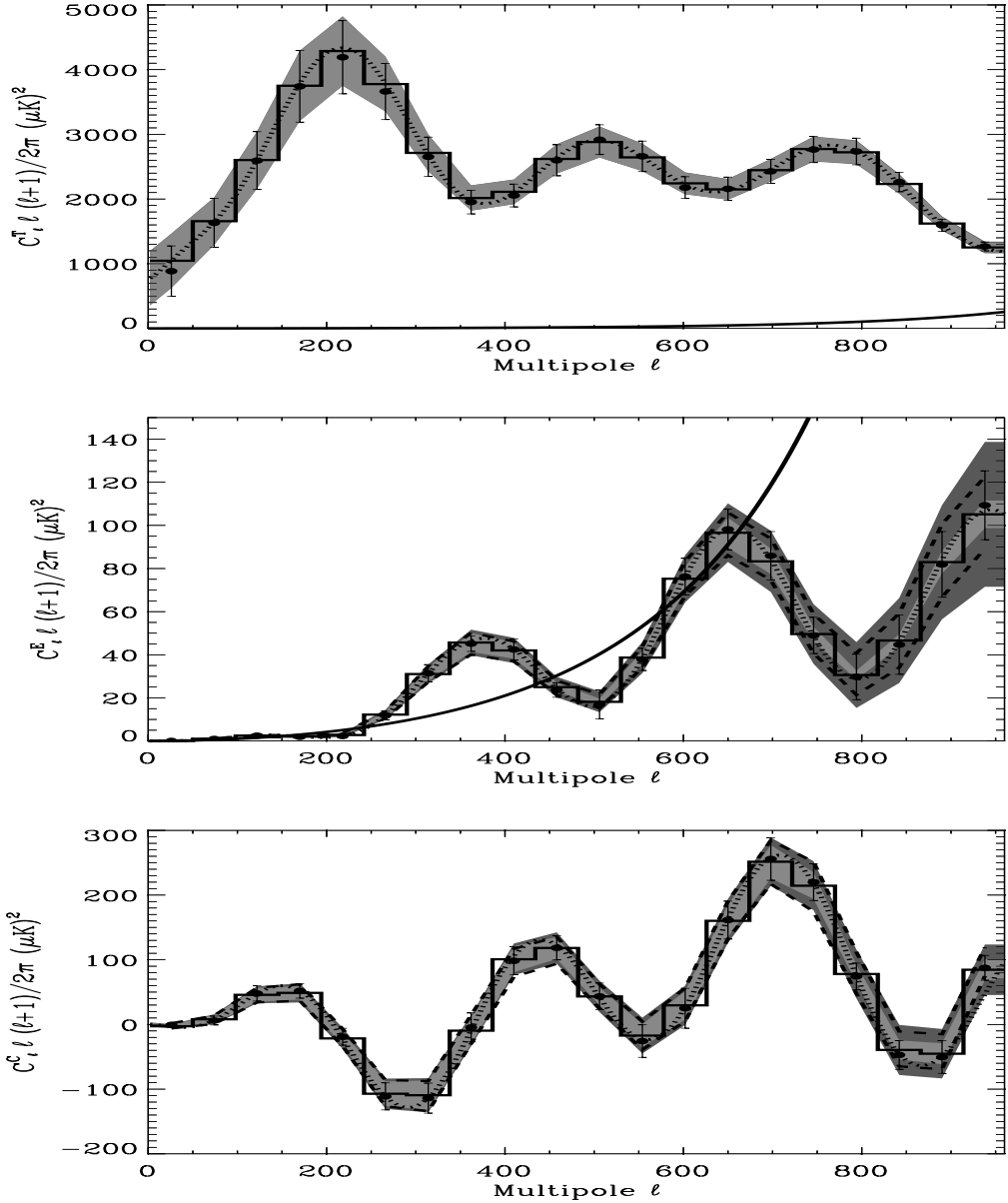,bbllx=0pt,bblly=0pt,bburx=696pt,bbury=842pt,height=18cm,width=22cm}
\caption{Same as figure (\ref{fig:polpcl}) but the dots here are the
average of 60 estimates from Monte Carlo
simulations. The error bars are the average deviations taken from the
simulations. The dotted line shows the average full sky spectrum. The
shaded areas which are plotted around the
binned full sky power spectrum (not shown) show the variance taken from
the approximate variance formula for uniform noise. The dashed lines show the
expected variance taken from the
inverse Fisher matrix.}
\label{fig:polpclnoavg}
\end{center}
\end{figure}

To test whether the method is bias or not, we did 60 Monte Carlo
simulations. The result of the average of these simulations is shown
in figure (\ref{fig:polpcl}). The method seems to be unbiased also for
the estimates of the polarisation power spectra. Note that the
expected noise variance taken
from the often used analytic formula for uniform noise (shaded areas
on the plot) given in \cite{master} and HGH here fails to predict the size of the error bars on the estimates. The
expected variance taken from the inverse Fisher matrix (dashed lines)
fits better with the error bars from Monte Carlo. The reason is that we
used a noise profile with increasing
noise from the centre of the disc and down to the edges, opposite of
the Gaussian window. This gives the observation with high $S/N$ higher significance in the analysis. This is similar to a result for the temperature power spectrum discussed further in HGH.\\

\section{Discussion}

We have presented a maximum likelihood method to simultaneously estimate the temperature and polarisation power spectra from high resolution CMB data in the presence of non-uniform noise and a symmetric Gabor window.
An extension of the power spectrum estimation method developed in
HGH has been made in order to estimate for
the polarisation power spectra in addition to the temperature power spectrum. In most standard theories for the early universe, the $B$ component polarisation will be too small to be observed by the MAP and Planck experiments. For this reason, the method has been tested here under the
assumption that the $B$ mode polarisation is negligible. In this case
the method appears to give unbiased estimates of the polarisation power
spectra also in the presence of non-uniform noise and a Gabor
window.\\

The kernels connecting the full sky polarisation power spectra with
the cut sky polarisation pseudo power spectra were studied and found
to be very similar to the kernel for the temperature power
spectrum. For this reason the effect of a cut sky and a Gabor window
on the polarisation power spectra is similar to the effect on the
temperature power spectrum. This explains that the method of estimating
the power spectrum from the pseudo power spectrum for polarisation was as successful as it was for temperature.\\

One issue which has not been studied fully here is the inclusion of the $B$
mode polarisation. We demonstrated that the $E$ and $B$ mode
polarisation power spectra are mixing on the cut sky making detections
of the much weaker $B$ component difficult. Further work needs to be
done in order to include the $B$ component in the likelihood analysis.\\

In HGH it was discussed how one can find the noise correlation matrix for temperature using Monte Carlo. This might be faster than the analytical approach presented here when the size of the dataset is very huge. With a sufficient number of Monte Carlo simulations this was shown to give similar error bars as the analytic treatment. The results for the temperature noise matrix is expected to be valid also for the polarisation noise matrices and can be used when the dataset is so big that the Monte-Carlo approach is significantly faster, or when correlated noise is present.\\

Another extension which was discussed in HGH was the simultaneous analysis of several patches on the CMB sky. This extension was shown to work for the temperature power spectrum and we expect that this could work also for polarisation, allowing data from several different experiments to be analysed together. In HGH it was shown that with this power spectrum estimation method, huge datasets like the ones to be expected from MAP or Planck can be analysed in a reasonable amount of time. The most time-consuming step in the method is the construction of the correlation matrix, which in the case of polarisation is 9 times longer. This makes also the joint temperature and polarisation power spectrum estimation feasible for huge datasets.

\section*{Acknowledgements}

We would like to thank E. Hivon, A. J. Banday and B. D. Wandelt for helpful discussions. We acknowledge the use of HEALPix \cite{healpix}
and CMBFAST \cite{cmbfast}. FKH was supported by a grant from the Norwegian Research Council.

\begin{appendix}

\section{Rotation Matrices}
\label{app:rotmat}
A spherical function $T(\mathbf{\hat n})$ is rotated by the operator $\hat
D(\alpha\beta\gamma)$ where $\alpha\beta\gamma$ are the three Euler
angles for rotations \cite{risbo} and the inverse rotation is
$\hat D(-\gamma-\beta-\alpha)$. For the spherical harmonic functions,
this operator takes the form,
\begin{equation}
Y_{\ell m}(\mathbf{\hat n}')=\sum_{m'=-\ell}^\ell
D_{m'm}^\ell(\alpha\beta\gamma)Y_{\ell m'}(\mathbf{\hat n}),
\end{equation}
where $D_{m'm}^\ell$ has the form
\begin{equation}
D_{m'm}^\ell(\alpha\beta\gamma)=e^{im'\alpha}d_{m'm}^\ell(\beta)e^{im\gamma}.
\end{equation}
Here $d_{m'm}^\ell(\beta)$ is a real coefficient with the following
property:
\begin{equation}
d_{m'm}^\ell(\beta)=d_{mm'}^\ell(-\beta).
\end{equation}
The D-functions also have the following property:
\begin{equation}
D_{m'm}^\ell(\alpha\beta\gamma)=\sum_{m''}D_{m'm''}^\ell(\alpha_2\beta_2\gamma_2)D_{m''m}^\ell(\alpha_1\beta_1\gamma_1),
\end{equation}
where $(\alpha\beta\gamma)$ is the result of the two consecutive
rotations $(\alpha_1\beta_1\gamma_1)$ and
$(\alpha_2\beta_2\gamma_2)$.

The complex conjugate of the rotation matrices can be written as
\begin{equation}
D^{\ell*}_{mm'}=(-1)^{m+m'}D^\ell_{(-m)(-m')}.
\end{equation}
See also Appendix (\ref{app:spinhar}).

\section{Spin-s Harmonics}
\label{app:spinhar}

The spherical harmonic functions $Y_{\ell m}(\mathbf{\hat n})$ can be
generalised to {\it spin-s harmonics} using the rotation matrices in
Appendix (\ref{app:rotmat}). The general definition is
\begin{equation}
D^\ell_{-sm}(\phi_2,\theta,\phi_1)=\sqrt{\frac{4\pi}{2\ell+1}}\
_sY_{\ell m}(\theta,\phi_2)e^{-is\phi_1},
\end{equation}
or in the form which will be mostly used in this paper
\begin{equation}
_sY_{\ell
m}(\theta,\phi)=\sqrt{\frac{2\ell+1}{4\pi}}D^\ell_{-sm}(\phi,\theta,0).
\end{equation}
The spin-s harmonics have the orthogonality and completeness relations
given by
\begin{eqnarray}
\int d\mathbf{\hat n}\ _sY_{\ell m}(\mathbf{\hat n})\ _sY_{\ell m}(\mathbf{\hat
n})&=&\delta_{\ell\ell'}\delta_{mm'}\\
\sum_{\ell m}\ _sY_{\ell m}(\mathbf{\hat n})\ _sY_{\ell m}(\mathbf{\hat
n}_0)&=&\delta(\mathbf{\hat n}-\mathbf{\hat n}_0).
\end{eqnarray}
The complex conjugate of the spin harmonics can be written
\begin{equation}
\label{eq:tennegm}
_sY_{\ell m}^*(\mathbf{\hat n})=(-1)^{s+m}\ _{-s}Y_{\ell(-m)}(\mathbf{\hat n}).
\end{equation}

\section{Some Wigner Symbol Relations}
\label{app:wig}

Throughout the paper, the Wigner 3j Symbols will be used
frequently. Here are some relations for these symbols, which are
used.
The orthogonality relation is,
\begin{equation}
\label{eq:wigort}
\sum_{mm'}\wigner{\ell}{\ell'}{\ell''}{m}{m'}{m''}\wigner{\ell}{\ell'}{L''}{m}{m'}{M''}=(2\ell''+1)^{-1}\delta_{\ell''L''}\delta_{m''M''}.
\end{equation}
The Wigner 3j Symbols can be represented as an integral of rotation
matrices (see Appendix(\ref{app:rotmat})),
\begin{equation}
\label{eq:wigd}
\frac{1}{8\pi^2}\int d\cos\theta d\phi d\gamma D_{m_1m'_1}^\ell
D_{m_2m'_2}^{\ell'}D_{m_3m'_3}^{\ell''}=\wigner{\ell}{\ell'}{\ell''}{m_1}{m_2}{m_3}\wigner{\ell}{\ell'}{\ell''}{m'_1}{m'_2}{m'_3}.
\end{equation}
This expression can be reduced to,
\begin{equation}
\label{eq:wigy}
\int d\mathbf{\hat n} Y_{\ell m}(\mathbf{\hat n})
Y_{\ell'm'}(\mathbf{\hat n})Y_{\ell''m''}(\mathbf{\hat n})=\sqrt{\frac{(2\ell+1)(2\ell'+1)(2\ell''+1)}{4\pi}}\wigner{\ell}{\ell'}{\ell''}{m}{m'}{m''}\wigo{}{}{}.
\end{equation}

\section{Extension of the Recurrence Relation to Polarisation}
\label{app:recpol}

For fast calculations of correlation matrices for polarisation, it
would be pleasant to have a recurrence relation for
\begin{equation}
h_2(\ell,\ell',m,m')\equiv\int d\mathbf{\hat n}G(\mathbf{\hat n})\ _2Y_{\ell m}(\mathbf{\hat n})\
_2Y_{\ell'm'}(\mathbf{\hat n}),
\end{equation}
similar to the one for $h(\ell,\ell',m,m')$ derived in the Appendix
of HGH. Again we simplify the notation by calling the function
$A_{\ell m}^{\ell'm'}$. Separating the spin-2 harmonic one can write
\begin{equation}
_2Y_{\ell m}(\theta,\phi)=\ _2\lambda_{\ell m}(\cos\theta)e^{-im\phi}.
\end{equation}
In this way one can write $A_{\ell m}^{\ell'm'}$ as
\begin{equation}
A_{\ell m}^{\ell'm'}=\int dx\ _2\lambda_{\ell m}(x)\
_2\lambda_{\ell'm'}(x)\underbrace{\int_{0}^{2\pi}d\phi
G(\theta,\phi)e^{-i\phi(m-m')}}_{\equiv F_{mm'}(x)},
\end{equation}
where $x=\cos\theta$. As before we define
\begin{equation}
X_{\ell m}^{\ell'm'}=\int dx x\ _2\lambda_{\ell m}(x)\
_2\lambda_{\ell'm'}(x) F_{mm'}(x),
\end{equation}
where obviously $F_{mm'}(x)^*=F_{m'm}(x)$ and therefore $X_{\ell
m}^{\ell'm'}=(X_{\ell'm'}^{\ell m})^*$. The next step is to use a
recurrence relation for spin-2 harmonics
\begin{equation}
x\ _2\lambda_{\ell m}(x)=p_{\ell m}\ _2\lambda_{(\ell+1)m}(x)-q_{\ell m}\
_2\lambda_{\ell m}(x)+p_{(\ell-1)m}\ _2\lambda_{(\ell-1)m}(x),
\end{equation}
where
\begin{eqnarray}
p_{\ell
m}&=&\frac{1}{\ell+1}\sqrt{\frac{((\ell+1)^2-m^2)((\ell+1)^2-4)}{4(\ell+1)^2-1}},\\
q_{\ell m}&=&\frac{2m}{\ell(\ell+1)}.
\end{eqnarray}
In this way one has
\begin{eqnarray}
\label{eq:x}
X_{\ell m}^{\ell'm'}&=&p_{\ell m}A_{(\ell+1)m}^{\ell'm'}-q_{\ell
m}A_{\ell m}^{\ell'm'}+p_{(\ell-1)m}A_{(\ell-1)m}^{\ell'm'},\\
\label{eq:xinv}
X_{\ell'm'}^{\ell m}&=&p_{\ell'm'}A_{(\ell'+1)m'}^{\ell m
}-q_{\ell'm'}A_{\ell'm'}^{\ell m}+p_{(\ell'-1)m'}A_{(\ell'-1)m'}^{\ell
m},
\end{eqnarray}
Subtracting the complex conjugate of equation (\ref{eq:xinv}) from
equation (\ref{eq:x}) the left side is zero and one is left with
 \begin{eqnarray}
A_{\ell m}^{\ell'm'}&=&\frac{1}{p_{(\ell'-1)m'}}\biggl[p_{\ell
m}A_{(\ell+1)m}^{(\ell'-1)m'}+(q_{(\ell'-1)m'}-q_{\ell m})A_{\ell
m}^{(\ell'-1)m'}\\
&&+p_{(\ell-1)m}A_{(\ell-1)m}^{(\ell'-1)m'}-p_{(\ell'-2)m'}A_{\ell m}^{(\ell'-2)m'}\biggr].
\end{eqnarray}
This is the final recursion formula.  The $A_{m' m'}^{\ell m}$ elements must
be provided before the recurrence is started. Then for each $(m,m')$, set
$\ell '=m'+1$ and let $\ell $ go from $\ell '$ and upwards, then set $\ell '=m'+2$ and again
let $\ell $ go from $\ell '$ and upwards. Continue to the desired size of $\ell '$.
Note that, in order to get all objects up to $A_{\ell _{max} m}^{\ell_{max} m'}$
one need to go up to $\ell =2\ell _{max}$ for each row of $\ell'$. This
is because of the $A_{(\ell+1) m}^{(\ell'-1) m'}$ term which demands
an object indexed $(\ell +1)$ in the previous $\ell '$ row.\\
To start the recurrence, one can precomputed the $A_{\ell m}^{m' m'}$ factors
fast and easily using FFT and a sum over rings on the grid. As an example, for
the HEALPix grid, we did it the following way,
\begin{equation}
A_{\ell m}^{m' m'}=\sum_r\lambda^r_{m'm'}\lambda^r_{\ell m}\sum_{j=0}^{N_r-1}
e^{-2\pi ij/N_r(m-m')}G_{rj},
\end{equation}
where the last part is the Fourier transform of the Gabor window,
calculated by FFT, $r$ is ring number on the grid and $j$ is azimuthal
position on each ring. Ring $r$ has $N_r$ pixels.

It turns out that the recurrence can be numerically unstable dependent
on the window and multipole, and in order to
avoid problems we (using double precision numbers) restart the
recurrence with a new set of precomputed $A^{\ell m}_{\ell' m'}$ for every
50th $\ell'$ row. However for some windows and multipoles the
recurrence can run for hundreds of $\ell$-rows without problems.\\\\

\section{The Polarisation Pseudo Power Spectrum}
\label{app:pseudopol}

The polarisation spherical harmonic
coefficients are
defined by means of the tensor spherical harmonics $_2 Y_{\ell
m}(\mathbf{\hat n})$ as 
\begin{eqnarray}
\label{eq:alm2}
a_{2,\ell m}&=&\int d\mathbf{\hat n} \ _2 Y^*_{\ell m}(\mathbf{\hat n})(Q+iU)(\mathbf{\hat n}),\\
\label{eq:almm2}
a_{-2,\ell m}&=&\int d\mathbf{\hat n} \ _{-2} Y^*_{\ell m}(\mathbf{\hat n})(Q-iU)(\mathbf{\hat n}),
\end{eqnarray}
and the inverse transforms are given as
\begin{eqnarray}
\label{eq:qu2}
(Q+iU)(\mathbf{\hat n})&=&\sum_{\ell' m'}a_{2,\ell' m'}\ _2 Y_{\ell'
m'}(\mathbf{\hat n})\\
(Q-iU)(\mathbf{\hat n})&=&\sum_{\ell' m'}a_{-2,\ell' m'}\ _{-2}
Y_{\ell' m'}(\mathbf{\hat n}).
\end{eqnarray}
It will be advantageous to write these spherical harmonics in terms of
the rotation matrices $D^\ell_{mm'}$ defined in Appendix
(\ref{app:rotmat}). Using the formulae in Appendix (\ref{app:spinhar})
one can write 
\begin{eqnarray}
\label{eq:y2d1}
\ _2 Y_{\ell
m}(\mathbf{\hat n})&=&\sqrt{\frac{2\ell+1}{4\pi}}D_{-2m}^\ell(\phi,\theta,0),\\
\label{eq:y2d2}
\
_{-2} Y_{\ell
m}(\mathbf{\hat n})&=&\sqrt{\frac{2\ell+1}{4\pi}}D_{2m}^\ell(\phi,\theta,0).\\
\end{eqnarray}
The corresponding complex conjugates can be written as (using the
relations in Appendix (\ref{app:rotmat}))
\begin{eqnarray}
\ _2Y_{\ell
m}^*(\mathbf{\hat
n})&=&\sqrt{\frac{2\ell+1}{4\pi}}D_{-2m}^{\ell*}(\phi,\theta,0)\\
\label{eq:y2d3}
&=&\sqrt{\frac{2\ell+1}{4\pi}}(-1)^mD_{2-m}^\ell(\phi,\theta,0),\\
\ _{-2}Y_{\ell
m}^*(\mathbf{\hat
n})&=&\sqrt{\frac{2\ell+1}{4\pi}}D_{2m}^{\ell*}(\phi,\theta,0)\\
\label{eq:y2d4}
&=&\sqrt{\frac{2\ell+1}{4\pi}}(-1)^mD_{-2-m}^\ell(\phi,\theta,0).
\end{eqnarray}
Finally, the power
spectrum can be written in terms of a `divergence free' $E$ component
and a `curl free' $B$ component
\begin{eqnarray}
a_{E,\ell m}&=&-\frac{1}{2}(a_{2,\ell m}+a_{-2,\ell m}),\\
a_{B,\ell m}&=&\frac{1}{2}i(a_{2,\ell m}-a_{-2,\ell m})
\end{eqnarray}\\

Now we will define the windowed coefficients $\tilde a_{\ell m}$ and
$\tilde C_\ell$ for polarisation in an analogous way as for
temperature. As in HGH we Legendre expand the Gabor
Window $G(\theta)$ which is an axissymmetric function centred at $\mathbf{\hat n}_0$,
\begin{eqnarray}
G(\theta)&=&\sum_{\ell''}\frac{2\ell''+1}{4\pi}g_{\ell''} P_{\ell''}(\cos\theta)=\sum_{\ell''
m''}g_{\ell''} Y_{\ell'' m''}(\mathbf{\hat n})Y^*_{\ell'' m''}(\mathbf{\hat n}_0).\\
&=&\sum_{\ell''m''}g_{\ell''}\sqrt{\frac{2\ell''+1}{4\pi}}D_{0 m''}^{\ell''}(\phi,\theta,0)Y_{\ell''m''}(\mathbf{\hat n}_0).
\end{eqnarray}

We define the windowed coefficients $\tilde a_{\ell m}$ as
\begin{equation}
\label{eq:pa2}
\tilde a_{2,\ell m}=\int d\mathbf{\hat n} \ _2 Y_{\ell m}^*(\mathbf{\hat n})(Q+iU)(\mathbf{\hat n})G(\mathbf{\hat n},\mathbf{\hat n}_0)
\end{equation}

Using the expression for (\ref{eq:qu2}) $(Q+iU)(\mathbf{\hat n})$ and
writing all $\ _2Y_{\ell m}$ as $D$-matrices using expressions
(\ref{eq:y2d1}), (\ref{eq:y2d2}), (\ref{eq:y2d3}) and (\ref{eq:y2d4}) one gets,
\begin{eqnarray}
\tilde a_{2,\ell
m}&=&\sum_{\ell'm'}a_{2,\ell'm'}\sum_{\ell''m''}g_{\ell''}\frac{\sqrt{(2\ell+1)(2\ell'+1)(2\ell''+1)}}{(4\pi)^{3/2}}(-1)^m\\
&\times&Y_{\ell''m''}(\mathbf{\hat n}_0)\int
d\mathbf{\hat n} D_{-2m'}^{\ell'}(\phi,\theta,0) D_{2-m}^\ell(\phi,\theta,0)
D_{0 m''}^{\ell''}(\phi,\theta,0)\\
&=&\sum_{\ell'm'}a_{2,\ell'm'}\sum_{\ell''m''}g_{\ell''}\frac{\sqrt{(2\ell+1)(2\ell'+1)(2\ell''+1)}}{(4\pi)^{3/2}}(-1)^m\frac{1}{2\pi}\\
&\times& Y_{\ell''m''}(\mathbf{\hat n}_0)\int
d\mathbf{\hat n} d\gamma D_{-2m'}^{\ell'}(\phi,\theta,\gamma)
D_{2-m}^\ell(\phi,\theta,\gamma) D_{0 m''}^{\ell''}(\phi,\theta,\gamma)\\ 
&=&\sum_{\ell'm'}a_{2,\ell'm'}h_2(\ell,\ell',m,m',\mathbf{\hat n}_0).
\end{eqnarray}

By using equation (\ref{eq:pa2}), one can also write this as,
\begin{equation}
\label{eq:h2intdef}
\tilde a_{2,\ell n}=\sum_{\ell'm'}a_{2,\ell'm'}\int d\mathbf{\hat n} G(\mathbf{\hat n},\mathbf{\hat n}_0)\
_2Y^*_{\ell m}(\mathbf{\hat n})\ _2Y_{\ell'm'}(\mathbf{\hat n}).
\end{equation}

Using the two last expressions, the $h_2$ function can be written in two ways (using relation
(\ref{eq:wigd}) for the last expression),
\begin{eqnarray}
h_2(\ell,\ell',m,m',\mathbf{\hat n}_0)&\equiv&\int d\mathbf{\hat n} G(\mathbf{\hat n},\mathbf{\hat n}_0)\
_2Y^*_{\ell m}(\mathbf{\hat n})\ _2Y_{\ell'm'}(\mathbf{\hat n})\\
&=&\sum_{\ell''m''}g_{\ell''}\sqrt{\frac{(2\ell+1)(2\ell'+1)(2\ell''+1)}{4\pi}}(-1)^m\\
&\times& Y_{\ell''m''}(\mathbf{\hat n}_0)\wigner{\ell}{\ell'}{\ell''}{-2}{2}{0}\wigner{\ell}{\ell'}{\ell''}{m'}{-m}{m''}.
\end{eqnarray}

As we soon will show, the polarisation pseudo power spectra are
rotationally invariant under rotation of the Gabor window. For that reason, one can put the centre of the
Gabor window on the north
pole giving,
\begin{equation}
\tilde a_{2,\ell m}=\sum_{\ell'}a_{2,\ell'm}h_2(\ell,\ell',m),
\end{equation}
where,
\begin{eqnarray}
h_2(\ell,\ell',m)&=&h_2(\ell,\ell',m,m,0)\\
&=&\sum_{\ell''}g_{\ell''}\frac{\sqrt{(2\ell+1)(2\ell'+1)}(2\ell''+1)}{4\pi}\\
&\times&(-1)^m\wigner{\ell}{\ell'}{\ell''}{-2}{2}{0}\wigner{\ell}{\ell'}{\ell''}{m}{-m}{0}.
\end{eqnarray}

Similarly one gets,
\begin{equation}
\tilde a_{-2,\ell m}=\sum_{\ell'}a_{-2,\ell'm}h_2(\ell,\ell',-m)
\end{equation}
and
\begin{eqnarray}
\tilde a_{E,\ell m}&=&-\frac{1}{2}(\tilde a_{2,\ell m}+\tilde
a_{-2,\ell m})\\
&=&\sum_{\ell'}a_{E,\ell'm}H_2(\ell,\ell',m)+i\sum_{\ell'}a_{B,\ell'm}H_{-2}(\ell,\ell',m)\\
\tilde a_{B,\ell m}&=&i\frac{1}{2}(\tilde a_{2,\ell m}-\tilde
a_{-2,\ell m})
\\&=&\sum_{\ell'}a_{B,\ell'm}H_2(\ell,\ell',m)-i\sum_{\ell'}a_{E,\ell'm}H_{-2}(\ell,\ell',m)
\end{eqnarray}

Please note that whereas $h(\ell,\ell',-m)=h(\ell,\ell',m)$ a similar
relation does not exist for $h_2(\ell,\ell',-m)$. Using the expression
above, one has that,

\begin{eqnarray}
h_2(\ell,\ell',-m)&=&\sum_{\ell''}g_{\ell''}\frac{\sqrt{(2\ell+1)(2\ell'+1)}(2\ell''+1)}{4\pi}(-1)^m\\
&\times&\wigner{\ell}{\ell'}{\ell''}{-2}{2}{0}\wigner{\ell}{\ell'}{\ell''}{m}{-m}{0}(-1)^{\ell+\ell'+\ell''}
\end{eqnarray}
The reason why this is not equal to $h_2(\ell,\ell',m)$ is that the
first Wigner symbol is not zero when $\ell+\ell'+\ell''$ is even,
which is the case when the whole lower row in the Wigner symbol is 0,
as in the case with $h(\ell,\ell',m)$. It is also obvious from the
expression (\ref{eq:h2intdef}). For the scalar case, the relation
$Y_{\ell(-m)}(\mathbf{\hat n})=(-1)^mY_{\ell m}(\mathbf{\hat n})$ ensures that
there is no dependency on $m$ in $h(\ell,\ell',m)$ whereas a similar
relation does not exist for the tensor harmonics (but see relation
(\ref{eq:tennegm})).\\

We have further defined,
\begin{eqnarray}
H_2(\ell,\ell',m)&=&\frac{1}{2}\left(h_2(\ell,\ell',m)+h_2(\ell,\ell',-m)\right)\\
H_{-2}(\ell,\ell',m)&=&\frac{1}{2}\left(h_2(\ell,\ell',m)-h_2(\ell,\ell',-m)\right)
\end{eqnarray}
which contrary to $h_2(\ell,\ell',m)$ have an $m$-symmetry
\begin{eqnarray}
H_2(\ell,\ell',-m)&=&H_2(\ell,\ell',m)\\
H_{-2}(\ell,\ell',-m)&=&-H_{-2}(\ell,\ell',m)
\end{eqnarray}

To find the $\tilde C_\ell$ and (later) the correlation matrices, the
following quantities will be needed
\begin{eqnarray}
\label{eq:aa1}
\VEV{\tilde a_{E,\ell m}\tilde
a_{E,\ell',m'}^*}&=&\delta_{mm'}\biggl[\sum_{\ell''}C_{\ell''}^EH_2(\ell,\ell'',m)H_2(\ell',\ell'',m)\\
&&+\sum_{\ell''}C_{\ell''}^BH_{-2}(\ell,\ell'',m)H_{-2}(\ell',\ell'',m)\biggr]\\
\VEV{\tilde a_{B,\ell m}\tilde
a_{B,\ell',m'}^*}&=&\delta_{mm'}\biggl[\sum_{\ell''}C_{\ell''}^BH_2(\ell,\ell'',m)H_2(\ell',\ell'',m)\\
&&+\sum_{\ell''}C_{\ell''}^EH_{-2}(\ell,\ell'',m)H_{-2}(\ell',\ell'',m)\biggr]\\
\VEV{\tilde a_{E,\ell m}\tilde
a_{B,\ell',m'}^*}&=&\delta_{mm'}i\biggl[\sum_{\ell''}C_{\ell''}^EH_2(\ell,\ell'',m)H_{-2}(\ell',\ell'',m)\\
&&+\sum_{\ell''}C_{\ell''}^BH_{-2}(\ell,\ell'',m)H_2(\ell',\ell'',m)\biggr]\\
\VEV{\tilde a_{E,\ell m}\tilde
a_{\ell',m'}^*}&=&\delta_{mm'}\sum_{\ell''}C_{\ell''}^CH_2(\ell,\ell'',m)h(\ell',\ell'',m)\\
\label{eq:aa2}
\VEV{\tilde a_{B,\ell m}\tilde
a_{\ell',m'}^*}&=&-i\delta_{mm'}\sum_{\ell''}C_{\ell''}^CH_{-2}(\ell,\ell'',m)h(\ell',\ell'',m)
\end{eqnarray}

For $\VEV{\tilde C_\ell^E}$ one now has,
\begin{equation}
\VEV{\tilde C_\ell^E}=\sum_m\frac{\VEV{\tilde a_{E,\ell m}^*\tilde a_{E,\ell
m}}}{2\ell+1}=\sum_{\ell'}C_{\ell'}^EK_2(\ell,\ell')+\sum_{\ell'}C_{\ell'}^BK_{-2}(\ell,\ell').
\end{equation}
Where,
\begin{equation}
K_{\pm2}(\ell,\ell')=\frac{1}{2\ell+1}\sum_mH_{\pm2}^2(\ell,\ell',m)
\end{equation}
Using the expression for $h_2(\ell,\ell',m)$, one gets
\begin{eqnarray}
K_{\pm2}(\ell,\ell')&=&\frac{1}{2}\frac{1}{2\ell+1}\sum_m\left(
h_2^2(\ell,\ell',m)\pm h_2(\ell,\ell',m)h_2(\ell,\ell',-m)\right)\\
&=&\sum_{\ell''L''}g_{\ell''}g_{L''}\frac{(2\ell'+1)(2\ell''+1)(2L''+1)}{32\pi^2}\\
&\times&\wigner{\ell}{\ell'}{\ell''}{-2}{2}{0}
\wigner{\ell}{\ell'}{L''}{-2}{2}{0}(1\pm(-1)^{\ell+\ell'+\ell''})\\
&\times&\underbrace{\sum_m\wigner{\ell}{\ell'}{\ell''}{m}{-m}{0}\wigner{\ell}{\ell'}{L''}{m}{-m}{0}}_{(2\ell''+1)^{-1}\delta_{\ell''L''}}\\
\label{eq:kpm2}
&=&\sum_{\ell''}g^2_{\ell''}\frac{(2\ell'+1)(2\ell''+1)}{32\pi^2}\wigner{\ell}{\ell'}{\ell''}{-2}{2}{0}^2\\
&\times&(1\pm(-1)^{\ell+\ell'+\ell''}).
\end{eqnarray}
Similarly,
\begin{eqnarray}
\VEV{\tilde C^B_\ell}&=&\sum_{\ell'}C_{\ell'}^B
K_2(\ell,\ell')+\sum_{\ell'}C_{\ell'}^EK_{-2}(\ell,\ell')\\
\label{eq:k0}
\VEV{\tilde C^C_\ell}&=&\sum_{\ell'}C_{\ell'}^C K_{20}(\ell,\ell')\\
K_{20}(\ell,\ell')&=&\frac{1}{2\ell+1}\sum_mH_2(\ell,\ell',m)h(\ell,\ell',m)\\
&=&\sum_{\ell''}g^2_{\ell''}\frac{(2\ell'+1)(2\ell''+1)}{16\pi^2}\wigner{\ell}{\ell'}{\ell''}{-2}{2}{0}\wigo{}{}{}.
\end{eqnarray}
Before studying these kernels we will first show the rotational invariance.
To show that $\VEV{\tilde C^E_\ell}$, $\VEV{\tilde C^B_\ell}$ and $\VEV{\tilde
C^C_\ell}$ are rotationally invariant under rotations of the window,
one can keeps the
dependency on the angle and the expression for the kernel turns out to
be,
\begin{eqnarray}
K_{\pm2}(\ell,\ell',\mathbf{\hat n}_0)&=&\frac{1}{2}\frac{1}{2\ell+1}\sum_{mm'}\biggl(
h_2^2(\ell,\ell',m,m',\mathbf{\hat n}_0)\\
&&\pm h_2(\ell,\ell',m,m',\mathbf{\hat n}_0)h_2(\ell,\ell',-m,-m',\mathbf{\hat n}_0)\biggr)\\
&=&\sum_{\ell''L''}\sum_{m''M''}g_{\ell''}g_{L''}\frac{(2\ell'+1)\sqrt{(2\ell''+1)(2L''+1)}}{8\pi}\\
&\times& Y_{\ell''m''}(\mathbf{\hat n}_0)Y_{L''M''}^*(\mathbf{\hat n}_0)\wigner{\ell}{\ell'}{\ell''}{-2}{2}{0}\wigner{\ell}{\ell'}{L''}{-2}{2}{0}\\
&\times&\underbrace{\sum_{mm'}\wigner{\ell}{\ell'}{\ell''}{m'}{-m}{m''}\wigner{\ell}{\ell'}{L''}{m'}{-m}{M''}}_{(2\ell''+1)^{-1}\delta_{\ell''L''}\delta_{m''M''}}\\
&\times&(1\pm(-1)^{\ell+\ell'+\ell''})\\
&=&\sum_{\ell''}g^2_{\ell''}\frac{2\ell'+1}{8\pi}\wigner{\ell}{\ell'}{\ell''}{-2}{2}{0}^2\underbrace{\sum_{m''}|Y_{\ell''m''}(\mathbf{\hat
n}_0)|^2}_{\frac{2\ell''+1}{4\pi}}\\
&\times&(1\pm(-1)^{\ell+\ell'+\ell''})\\
&&=K_{\pm2}(\ell,\ell').
\end{eqnarray}
This is independent on the angle $\mathbf{\hat n}_0$ which shows the rotational
invariance of the pseudo polarisation power spectra.\\

\section{Rotational Invariance}
\label{app:rotinv}

We now want to show that the pseudo power spectra for polarisation are
(as the temperature power spectra) rotationally invariant under a
common rotation of the sky and window. First, note that the
rotation matrices $D^\ell_{mm'}$ are rotating both the normal
spherical harmonics and the spin-s harmonics. This is easy to
show. Assume that one wants to rotate $\ _sY_{\ell m}(\mathbf{\hat n})$ with the
Euler angles $(\alpha,\beta,\gamma)$. Using the formula for the normal
spherical harmonics one gets,
\begin{eqnarray}
\ _sY_{\ell m}^\textrm{rot}(\mathbf{\hat n})&=&\sum_{m'}D_{m'm}^\ell(\alpha,\beta,\gamma)\ _sY_{\ell
m'}(\mathbf{\hat n})\\
&=&\sqrt{\frac{2\ell+1}{4\pi}}\sum_{m'}D_{m'm}^\ell(\alpha,\beta,\gamma)D_{-sm'}^\ell(\phi,\theta,0)\\
&=&D_{-sm}^\ell(\phi_{rot},\theta_{rot},0)\sqrt{\frac{2\ell+1}{4\pi}}\\
&=&\ _sY_{\ell m}(\mathbf{\hat n}_{rot}),
\end{eqnarray}
where $\mathbf{\hat n}_{rot}$ is the rotation of the angle $\mathbf{\hat n}$ by
$(\alpha,\beta,\gamma)$. This is clearly general for all
spin-s harmonics. Therefore we use the method from HGH to
show that polarisation pseudo power spectra are rotationally invariant,

Consider a rotation of the sky and window by the angles
$(-\gamma-\beta-\alpha)$. Then the $\tilde a_{s,\ell m}$ becomes,
\begin{equation}
\tilde a_{s,\ell m}^{\textrm{rot}}=\int d\mathbf{\hat n}[\hat
D(-\gamma-\beta-\alpha) T(\mathbf{\hat n}) G(\mathbf{\hat n})] \ _sY^*_{\ell m}(\mathbf{\hat n}).
\end{equation}
If one makes the inverse rotation of the integration angle $\mathbf{\hat n}$, one
can write this as;
\begin{equation}
\tilde a_{s,\ell m}^{\textrm{rot}}=\int d\mathbf{\hat n} T(\mathbf{\hat n})G(\mathbf{\hat n})[\hat
D^*(\alpha\beta\gamma)\ _sY^*_{\ell m}(\mathbf{\hat n})],
\end{equation}
which is just
\begin{equation}
\tilde a_{s,\ell
m}^{\textrm{rot}}=\sum_{m'}D_{m'm}^{\ell*}(\alpha\beta\gamma)\int
T(\mathbf{\hat n})G(\mathbf{\hat n})\ _sY^*_{\ell m'}(\mathbf{\hat n}).
\end{equation}
The last integral can be identified as the normal $\tilde a_{s,\ell m}$.
\begin{equation}
\tilde a_{s,\ell
m}^{\textrm{rot}}=\sum_{m'}D_{m'm}^{\ell*}(\alpha\beta\gamma)\tilde
a_{s,\ell m}.
\end{equation}
Thus,
\begin{equation}
\tilde a_{E,\ell
m}^{\textrm{rot}}=\sum_{m'}D_{m'm}^{\ell*}(\alpha\beta\gamma)\tilde
a_{E,\ell m}\ \ \ \tilde a_{B,\ell
m}^{\textrm{rot}}=\sum_{m'}D_{m'm}^{\ell*}(\alpha\beta\gamma)\tilde
a_{B,\ell m}.
\end{equation}
For $\tilde C_{E,\ell m}$ (and analogously for $\tilde C_{B,\ell m}$
and $\tilde C_{C,\ell m}$) one gets
\begin{eqnarray}
\tilde C_{E,\ell}^\textrm{rot}&=&\frac{1}{2\ell+1}\sum_m a^\textrm{rot}_{E,\ell
m}a^\textrm{rot*}_{E,\ell m}\\
&=&\frac{1}{2\ell+1}\sum_m\sum_{m'}\sum_{m''}D_{m'm}^\ell(\alpha\beta\gamma)D_{m''m}^{\ell*}(\alpha\beta\gamma)\tilde
a_{E,\ell m'} \tilde a_{E,\ell m''}^*\\
&=&\frac{1}{2\ell+1}\sum_{m'm''}\tilde
a_{E,\ell m'} \tilde a_{E,\ell m''}^*\underbrace{\sum_m
D_{m'm}^\ell(\alpha\beta\gamma)D_{m''m}^{\ell*}(\alpha\beta\gamma)}_{\delta_{mm'}}\\
&=&\tilde C_{E,\ell}.
\end{eqnarray}

\section{The Polarisation Correlation Matrix}
\label{app:polcormat}

To find the correlation matrix $\mt{M}$ for likelihood estimation of
the polarisation power spectra one needs the formulae given in
equations (\ref{eq:aa1}) to (\ref{eq:aa2}). As shown there the
correlations of the pseudo $\tilde a_{\ell m}$ coefficients can be written in terms of the
$h(\ell,\ell',m)$ function from HGH and the
$H_2(\ell,\ell',m)$ and $H_{-2}(\ell,\ell',m)$ functions. These
function can be quickly calculated using the important recursion
formulae deduced in Appendix (\ref{app:recpol}) and in the Appendix of HGH. The
starting points of these recursions can also be quickly provided using
summations and FFT as explained in HGH. We will
now show that the correlation function $\mt{M}$ can be expressed in
terms of these functions and for this reason can be calculated
quickly.\\

The pseudo power spectra can be written as
\begin{eqnarray}
\VEV{\tilde C^T_\ell}&=&\sum_m\frac{\VEV{\tilde a^T_{\ell m}\tilde a^{T*}_{\ell m}}}{2\ell+1},\\
\VEV{\tilde C^E_\ell}&=&\sum_m\frac{\VEV{\tilde a^E_{\ell m}\tilde a^{E*}_{\ell m}}}{2\ell+1},\\
\VEV{\tilde C^C_\ell}&=&\sum_m\frac{\VEV{\tilde a^T_{\ell m}\tilde a^{E*}_{\ell m}}}{2\ell+1}.
\end{eqnarray}

To find the correlation function between $\tilde C_\ell$ for
polarisation one can follow the
same steps as for the temperature correlation functions in HGH, and get,
\begin{eqnarray}
M_{EE,\ell\ell'}&=&\frac{2}{(2\ell+1)(2\ell'+1)}\sum_m\VEV{\tilde a_{E,\ell
m}\tilde a_{E,\ell' m}^*}^2,\\
M_{BB,\ell\ell'}&=&\frac{2}{(2\ell+1)(2\ell'+1)}\sum_m\VEV{\tilde a_{B,\ell
m}\tilde a_{B,\ell' m}^*}^2,\\
M_{CC,\ell\ell'}&=&\frac{1}{(2\ell+1)(2\ell'+1)}\sum_m\biggl[\VEV{\tilde a_{E,\ell
m}\tilde a_{E,\ell' m}^*}\VEV{\tilde a_{\ell
m}\tilde a_{\ell' m}^*}\\
&&+\VEV{\tilde a_{E,\ell
m}\tilde a_{\ell' m}^*}\VEV{\tilde a_{E,\ell'
m}\tilde a_{\ell m}^*}\biggr],\\
M_{ET,\ell\ell'}&=&\frac{2}{(2\ell+1)(2\ell'+1)}\sum_m\VEV{\tilde a_{E,\ell
m}\tilde a_{\ell' m}^*}^2,\\
M_{BT,\ell\ell'}&=&\frac{2}{(2\ell+1)(2\ell'+1)}\sum_m|\VEV{\tilde a_{B,\ell
m}\tilde a_{\ell' m}^*}|^2,\\
M_{CT,\ell\ell'}&=&\frac{2}{(2\ell+1)(2\ell'+1)}\sum_m\VEV{\tilde a_{E,\ell
m}\tilde a_{\ell' m}^*}\VEV{\tilde a_{\ell
m}\tilde a_{\ell' m}^*},\\
M_{EB,\ell\ell'}&=&\frac{2}{(2\ell+1)(2\ell'+1)}\sum_m|\VEV{\tilde a_{E,\ell
m}\tilde a_{B,\ell' m}^*}|^2,\\
M_{EC,\ell\ell'}&=&\frac{2}{(2\ell+1)(2\ell'+1)}\sum_m\VEV{\tilde a_{E,\ell
m}\tilde a_{E,\ell' m}^*}\VEV{\tilde a_{E,\ell
m}\tilde a_{\ell' m}^*},\\
M_{BC,\ell\ell'}&=&\frac{2}{(2\ell+1)(2\ell'+1)}\sum_m\VEV{\tilde a_{E,\ell'
m}\tilde a_{B,\ell m}^*}\VEV{\tilde a_{\ell'
m}^*\tilde a_{B,\ell m}},\\
\end{eqnarray}
where the correlation between $\tilde a_{\ell m}$ are given in
equations (\ref{eq:aa1}) to (\ref{eq:aa2}) as sums of
$h(\ell,\ell',m)$ and $H_{\pm2}(\ell,\ell',m)$.

\section{Polarisation with Noise}
\label{app:polnoise}

Analogously to HGH we will now discuss the
noise pseudo power spectra and the noise correlation matrix for polarisation. Each
pixel in the temperature map has a noise temperature $n_j$ and for the
polarisation maps we assume  $n^Q_j$ and $n^U_j$ to have the following
properties,
\begin{eqnarray}
\VEV{n_j}=0,&\VEV{n_jn_{j'}}=\delta_{jj'}(\sigma^T_j)^2,&\\
\VEV{n^Q_j}=0,&\VEV{n^Q_jn^Q_j}=\delta_{jj'}(\sigma^P_j)^2,\\
\VEV{n^U_i}=0,&\VEV{n^U_in^U_j}=\delta_{ij}(\sigma^P_j)^2,
\end{eqnarray}
We also assume that there is no correlation between noise in the
different maps T,Q and U.
For the full sky one has,
\begin{eqnarray}
\VEV{a^N_{\ell m}a_{\ell m}^{N*}}&=&\sum_j(\sigma^T_j)^2|Y^j_{\ell m}|^2\\
\VEV{a^N_{2,\ell m}a_{2,\ell m}^{N*}}&=&2\sum_j(\sigma^P_j)^2|\ _2Y_{\ell m}^j|^2,\\
\VEV{a^N_{-2,\ell m}a_{-2,\ell m}^{N*}}&=&2\sum_j(\sigma^P_j)^2|\ _{-2}Y_{\ell m}^j|^2,\\
\VEV{a^N_{2,\ell m}a_{-2,\ell m}^{N*}}&=&0,\\
\VEV{a^N_{E,\ell m}a_{E,\ell
m}^{N*}}&=&2\sum_j(\sigma^P_j)^2(|\ _2Y_{\ell m}^j|^2+|\ _{-2}Y_{\ell m}^j|^2|),\\
\VEV{a^N_{B,\ell m}a_{B,\ell
m}^{N*}}&=&\VEV{a^N_{E,\ell m}a_{E,\ell
m}^{N*}},\\
\VEV{a^N_{E,\ell m}a_{B,\ell
m}^{N*}}&=&0,
\end{eqnarray}
which for this type of noise gives $C^{EN}_\ell=C^{BN}_\ell$.\\

The pseudo $a_{2,\ell m}$ coefficients can now be found using
equations (\ref{eq:alm2}) and (\ref{eq:almm2}) we define
\begin{equation}
\tilde a^N_{\pm2,\ell m}=\sum_j(n_j^Q\pm in_j^U)G_j\ _{\pm2}Y_{\ell
m}^j,
\end{equation}
for an axissymmetric Gabor window $G$ having the value $G_j$ in
pixel $j$.
The $E$ and $B$ components are then similarly
\begin{eqnarray}
\tilde a^N_{E,\ell m}&=&-\frac{1}{2}\sum_j\biggl[n_j^Q\left(\ _2Y_{\ell
m}^j+\ _{-2}Y_{\ell m}^j\right)+in_j^U\left(\ _2Y_{\ell m}^j-\
_{-2}Y_{\ell m}^j\right)\biggr]G_j,\\
\tilde a^N_{B,\ell m}&=&\frac{1}{2}i\sum_j\biggl[n_j^Q\left(\ _2Y_{\ell
m}^j-\ _{-2}Y_{\ell m}^j\right)+in_j^U\left(\ _2Y_{\ell m}^j+\
_{-2}Y_{\ell m}^j\right)\biggr]G_j.\\
\end{eqnarray}
The correlations between these coefficients are 
\begin{eqnarray}
\VEV{\tilde a^N_{E,\ell m}\tilde
a^{N*}_{E,\ell'm'}}&=&\frac{1}{4}\sum_j\biggl[(\sigma^P_j)^2\left(\
_2Y_{\ell m}^j+\ _{-2}Y_{\ell m}^j\right)\\
&&\times\left(\ _2Y_{\ell' m'}^j+\ _{-2}Y_{\ell' m'}^j\right)\\
&&+(\sigma^P_j)^2\left(\ _2Y_{\ell m}^j-\ _{-2}Y_{\ell
m}^j\right)\\
&&\times\left(\ _2Y_{\ell' m'}^j-\ _{-2}Y_{\ell'
m'}^j\right)\biggr]G_j^2\\
\label{eq:eeh2}
&=&\frac{1}{2}[h_2'(\ell,\ell',m,m')\\
\nonumber&&+(-1)^{m+m'}h'_2(\ell,\ell',-m,-m')]\\
&\equiv&H'_2(\ell,\ell',m,m'),
\end{eqnarray}
where the last line defines $H'_2(\ell,\ell',m,m')$. The
$h'_2(\ell,\ell',m,m')$ is defined similar to the
$h'(\ell,\ell',m,m')$ function in HGH
\begin{equation}
h'_2(\ell,\ell',m,m')=\sum_jG_j^2(\sigma^P_j)^2\ _2Y_{\ell
m}^j\ _2Y_{\ell'm'}^j,
\end{equation}
Note the following relation which was used to obtain equation
(\ref{eq:eeh2})
\begin{equation}
\sum_jG_j^2(\sigma^P_j)^2\ _{-2}Y_{\ell m}^j\
_{-2}Y_{\ell'm'}^j=(-1)^{m+m'}h_2(\ell,\ell',-m,-m').
\end{equation}
Again, one can see that when the Gabor window AND noise have azimuthal
symmetry this reduces simply to,
\begin{equation}
h'_2(\ell,\ell',m,m')=h'_2(\ell,\ell',m).
\end{equation}

In a similar manner the other $a_{\ell m}$ relations can be found
\begin{eqnarray}
\VEV{\tilde a^N_{B,\ell m}\tilde a^{N*}_{B,\ell'm'}}&=&\VEV{\tilde a^N_{E,\ell
m}\tilde a^{N*}_{E,\ell'm'}}\\
\VEV{\tilde a^N_{E,\ell m}\tilde
a^{N*}_{B,\ell'm'}}&=&\frac{1}{2}i[h'_2(\ell,\ell',m,m')-\\
&&\nonumber(-1)^{m+m'}h'_2(\ell,\ell',-m,-m')]\\
&\equiv&iH'_{-2}(\ell,\ell',m,m'),
\end{eqnarray}
where the last line again defines $H'_{-2}(\ell,\ell',m,m')$. The
$H_{\pm2}(\ell,\ell',m,m')$ functions which are needed to find the noise
correlation matrices can be quickly calculated using the recursion in
appendix (\ref{app:recpol}).\\

Using these relations one can now find the polarisation pseudo spectra
\begin{eqnarray}
\VEV{\tilde C_\ell^{EN}}=\VEV{\tilde C_\ell^{BN}}&=&\frac{1}{2\ell+1}\sum_m\VEV{\tilde
a_{E,\ell m}^N\tilde a_{E,\ell m}^{N*}},\\
\VEV{\tilde C_\ell^{CN}}&=&0
\end{eqnarray}

One can further use this to find the noise correlation matrices
$M_{ZZ',\ell\ell'}^N$, defined as,
\begin{equation}
M_{ZZ',\ell\ell'}^N=\VEV{C_\ell^{ZN}C_{\ell'}^{Z'N}}-\VEV{C_\ell^{ZN}}\VEV{C_{\ell'}^{Z'N}},
\end{equation}
where $Z=\{T,E,B,C\}$. We find,
\begin{eqnarray}
M_{TT,\ell\ell'}^N&=&\frac{2}{2\ell+1}\sum_{mm'}|\VEV{\tilde a_{\ell
m}^N\tilde a_{\ell' m'}^{N*}}|^2,\\
M_{EE,\ell\ell'}^N&=&\frac{2}{2\ell+1}\sum_{mm'}|\VEV{\tilde a_{E,\ell
m}^N\tilde a_{E,\ell' m'}^{N*}}|^2,\\
M_{BB,\ell\ell'}^N&=&\frac{2}{2\ell+1}\sum_{mm'}|\VEV{\tilde a_{B,\ell
m}^N\tilde a_{B,\ell' m'}^{N*}}|^2,\\
M_{CC,\ell\ell'}^N&=&\frac{1}{2\ell+1}\sum_{mm'}\VEV{\tilde a_{E,\ell
m}^N\tilde a_{E,\ell' m}^{N*}}\VEV{\tilde a_{\ell
m}^N\tilde a_{\ell' m'}^{N*}},\\
M_{EB,\ell\ell'}^N&=&\frac{2}{2\ell+1}\sum_{mm'}|\VEV{\tilde a_{E,\ell
m}^N\tilde a_{B,\ell' m'}^{N*}}|^2,\\
\end{eqnarray}
all others combinations are zero. We then find the total correlation
matrix $M_{ZZ',\ell\ell'}$ consisting of both signal and noise. As for
temperature, this is not simply the sum of the correlation matrix for
signal and noise, one also gets cross terms. The final result is
\begin{eqnarray}
M_{TT,\ell\ell'}&=&M_{TT,\ell\ell'}^S+M_{TT,\ell\ell'}^N+\\
&&\frac{4}{2\ell+1}\sum_{mm'}\VEV{\tilde
a_{\ell m}^S\tilde a_{\ell'm'}^{S*}}\VEV{\tilde
a_{\ell m}^N\tilde a_{\ell'm'}^{N*}},\\
M_{EE,\ell\ell'}&=&M_{EE,\ell\ell'}^S+M_{EE,\ell\ell'}^N+\\
&&\frac{4}{2\ell+1}\sum_{mm'}\VEV{\tilde
a_{E,\ell m}^S\tilde a_{E,\ell'm'}^{S*}}\VEV{\tilde
a_{E,\ell m}^N\tilde a_{E,\ell'm'}^{N*}},\\
M_{BB,\ell\ell'}&=&M_{BB,\ell\ell'}^S+M_{BB,\ell\ell'}^N+\\
&&\frac{4}{2\ell+1}\sum_{mm'}\VEV{\tilde
a_{B,\ell m}^S\tilde a_{B,\ell'm'}^{S*}}\VEV{\tilde
a_{B,\ell m}^N\tilde a_{B,\ell'm'}^{N*}},\\
M_{CC,\ell\ell'}&=&M_{CC,\ell\ell'}^S+M_{CC,\ell\ell'}^N+\\
&&\frac{1}{2\ell+1}\sum_{mm'}\biggl(\VEV{\tilde
a_{E,\ell m}^S\tilde a_{E,\ell'm'}^{S*}}\VEV{\tilde
a_{\ell m}^N\tilde a_{\ell'm'}^{N*}}\\
&&+\VEV{\tilde
a_{\ell m}^S\tilde a_{\ell'm'}^{S*}}\VEV{\tilde
a_{E,\ell m}^N\tilde a_{E,\ell'm'}^{N*}}\biggr),\\
M_{TE,\ell\ell'}&=&M_{TE,\ell\ell'}^S\\
M_{TB,\ell\ell'}&=&M_{TB,\ell\ell'}^S\\
M_{TC,\ell\ell'}&=&M_{TC,\ell\ell'}^S+\frac{2}{2\ell+1}\sum_{mm'}\VEV{\tilde
a_{\ell m}^S\tilde a_{E,\ell'm'}^{S*}}\VEV{\tilde
a_{\ell m}^N\tilde a_{\ell'm'}^{N*}},\\
M_{EB,\ell\ell'}&=&M_{EB,\ell\ell'}^S+M_{EB,\ell\ell'}^N+\\
&&\frac{4}{2\ell+1}\sum_{mm'}\VEV{\tilde
a_{E,\ell m}^S\tilde a_{B,\ell'm'}^{S*}}\VEV{\tilde
a_{E,\ell m}^N\tilde a_{B,\ell'm'}^{N*}},\\
M_{EC,\ell\ell'}&=&M_{EC,\ell\ell'}^S+\frac{2}{2\ell+1}\sum_{mm'}\VEV{\tilde
a_{E,\ell m}^S\tilde a_{\ell'm'}^{S*}}\VEV{\tilde
a_{E,\ell m}^N\tilde a_{E,\ell'm'}^{N*}},\\
M_{BC,\ell\ell'}&=&M_{BC,\ell\ell'}^S+\frac{2}{2\ell+1}\sum_{mm'}\VEV{\tilde
a_{B,\ell m}^S\tilde a_{\ell'm'}^{S*}}\VEV{\tilde
a_{B,\ell m}^N\tilde a_{E,\ell'm'}^{N*}},
\end{eqnarray}

\end{appendix}

\end{document}